\documentclass[a4paper,14pt]{book}
\usepackage{epsfig}
\usepackage{subfigure}
\usepackage{amsmath}   
\usepackage{amssymb}
\usepackage[spanish]{babel}
\usepackage[latin1]{inputenc}
\usepackage{newcent}
\usepackage{scalefnt}
\usepackage{extsizes} 
\usepackage{fancyhdr} 

\newcommand{\da}{^\dagger}
\newcommand{\bra}[1]{\left<#1\right|}
\newcommand{\ket}[1]{\left|#1\right>}
\newcommand{\braketo}[1]{\langle #1 \rangle}
\newcommand{\abs}[1]{\bigl| #1 \bigr|}

\newcommand{\su}{\uparrow}
\newcommand{\sd}{\downarrow}

\begin{document}
\frontmatter
\title{\vskip -2cm\textbf{Tesis de Licenciatura}\\ \vskip 4cm Generando Entrelazamiento en una cadena XY}
\author{Christian Tomás Schmiegelow\\ \\
\textbf{Director}\\
Raúl Dante Rossignoli\\\\\\\\
\small Presentada 30/06/2006\\
\small Última Revisión 22/12/2006
\\\\\\\\
\small UNIVERSIDAD NACIONAL DE LA PLATA\\
\small FACULTAD DE CIENCIAS EXACTAS\\
\small DEPARTAMENTO DE FÍSICA}
\date{}

\maketitle

\chapter*{Agradecimientos}
\begin{description}
\item[$\diamondsuit$]A Mamá, Pato y Clara. Por haberme acompañado con amor en este camino como lo hubieran hecho en cualquier otro que hubiera elegido.
\item[$\Im$]A Papá.
\item[$\clubsuit$]A mis Amigos. Por ser un inconmesurable apoyo emocional. En espacial a Mateo con quien aprendí mucha de la física que sé, tratando de explicársela como si fuese un niño.
\item[$\heartsuit$]A Maik.
\item[$\bigstar$]A todos mis Compañeros. No sólo porque no podría haber aprendido física sólo, sino porque son tambíen mis amigos. En especial a Pablo con quien cursé los primeros años, a Patricia por su visón escéptica de la física, a Carlos y a Mauricio con quienes cursé casi toda la carrera y al Zeke por ser tan insoportablemente Zeke.
\item[$\odot$]A mis compañeros de militancia de El Pelo de Einstein.
\item[$\Join$]A mis compañeros de casa Nati y Zaki.
\item[$\wp$]A mis Profesores. Tanto a tres profesores muy especiales del secundario que me guiaron hacia aquí, como los docentes-investigadores de esta casa que fueron una guia incondicional desde que ingresé a la carrera en el 2001.
\item[$\Re$]A Raul Rossignoli. Por su calma cada vez que me equivoqué, por los dolores de cabeza que le debo haber dado, por su apoyo a pesar de mis idas y vueltas, por haber dictado la materia optativa ``Intro. a la Comp. Cuántica y Teoría de la Información'' y por el infinito esfuerzo revisando y corrigiendo errores en este trabajo.
\end{description}

\tableofcontents

\chapter*{Introducción}
\addcontentsline{toc}{chapter}{Introducción}
Desde 1935, cuando  Einstein, Podolsky y Rosen presentaron un trabajo cuestionando la completitud de la mecánica cuántica, se reconoce que el entrelazamiento de sistemas cuánticos es una de las propiedades más intrigantes de la mecánica cuántica. En los últimos 20 años se han desarrollado técnicas, tanto experimentales como teóricas, que permiten utilizar esta propiedad como recurso para resolver problemas computacionales con eficiencias muy superiores a las que clásicamente se  habian demostrado  insuperables. En años recientes se han logrado también avances muy importantes en la comprensión y clasificación del entrelazamiento cuántico, aunque subsisten aun muchos puntos importantes por resolver e investigar.\\

Estudio en este trabajo la capacidad de generar entrelazamiento de una cadena de espines con acoplamiento de Heisenberg XY y un campo magnético uniforme a partir de un estado inicial en el que los espines están completamente alineados. Se encuentra que la capacidad de generar estados entrelazados no muestra un comportamiento monótono con el campo presentando, en cambio, ``plateaus'' y resonancias.  También se muestra que, a pesar de que la anisotropía es necesaria para que  se generen estados entrelazados, una mayor anisotropía no implica necesariamente mejores condiciones para generar entrelazamiento que sirva para usarse en una computadora cuántica.\\

Este trabajo está dividido en tres partes. En la primera se exponen nociones preliminares sobre entrelazamiento y computación cuántica necesarias para comprender el estudio específico de la cadena XY que se realiza en la segunda. La primeras dos partes están divididas en dos capítulos cada una.  En  el Capítulo 1 se da una introducción al concepto de entrelazamiento,  una noción del formalismo de la computación cuántica y finalmente se discute la relación entre estos dos. En el Capítulo 2 se presentan y discuten distintas medidas de entrelazamiento. La segunda parte empieza en el Capítulo 3 donde  se presenta la cadena XY como una posible realización de una computadora cuántica, se comenta sobre las simetrías relevantes, se lo resuelve para dos casos simples y se muestra la  capacidad de este sistema para generar entrelazamiento en estos casos. En el Capítulo 4 se resuelve la evolución temporal de la cadena XY en un caso mucho más general y se realizan diversas aproximaciones que permiten describir de manera simple su capacidad para generar entrelazamiento en distintos casos. Finalmente en la tercer parte se incluyen cuatro apéndices con el objetivo de ampliar algunos temas relevantes.

Los resultados obenidos en este trabajo han sido aceptados para su publicación en la revista \emph{Physical Review A} en un trabajo conjunto con Raul Rossignoli. En la publicación aprarecen además resultados que incluyen: cadenas con cantidad par e impar de qubits, resonancia de pares entrelazados de paridad positiva, un estudio detallado del comportamiento para $v=g$ y $b=0$ donde la evolución es estrictamente periódica y resultados de diagonalización directa para $n=4$.
 
\mainmatter
\part{Nociones Preliminares}
\chapter[Entrelazamiento y Computación Cuántica]{Entrelazamiento y Computación Cuántica}
La idea de hacer máquinas que calculen que se basen en las reglas de la Mecánica cuántica muchas veces se le atribuye a Richard Feynman. Sin embargo no fue hasta algunos años después de su muerte que,  basados en los trabajos de Deutsch, independientemente Shor y Grover descubren algoritmos que permiten resolver más eficientemente algunos problemas que una computadora clásica. Estos trabajos fueron el puntapié inicial que llevó a armar un esquema en el que la Computación Cuántica resulta una opción viable. La computación cuántica explota un aspecto fundamental de la mecánica cuántica: el \emph{entrelazamiento}.\\

\section[Entrelazamiento y Correlaciones Clásicas]{Entrelazamiento y Correlaciones \newline Clásicas}
El día a día del mundo que percibimos puede ser bien descripto mediante la física clásica. En él observamos continuamente correlaciones entre eventos: el FMI impone, por medio del Gobierno local, una nefasta ley de educación superior - los estudiantes manifiestan, los estudiantes manifiestan - sale la policía a reprimir, etc; o pongo un disco de \emph{Pixies} en el minicomponente - escucho \emph{Where is my Mind}. De todos estos eventos hay algunos que está más correlacionados que otros. Tomemos el ejemplo del disco de \emph{Pixies}. Si tengo 3 álbumes de esta banda pero en sólo uno está ese tema entonces hay una correlación de 1/3 de escuchar el tema (suponiendo que el equipo anda bien y ninguna otra cosa sale mal). Si en cambio estoy escuchando \emph{Where is my Mind} hay una correlación máxima, de valor 1, de que el disco que puse en el minicomponente sea el de \emph{Pixies} y no otro porque sé que ese tema no lo tengo en ningún otro disco. Dos eventos clásicos están máximamente correlacionados cuando la existencia de uno implica la del otro y viceversa. Clásicamente este límite no puede superarse.

La mecánica cuántica es un conjunto de reglas que fueron desarrolladas para describir el comportamiento de objetos microscópicos que, como decía Heisenberg, funciona; es decir describe adecuadamete la realidad. Desde sus comienzos y hasta hoy en día la interpretación de estas reglas es tema de discusión. La interpretación de Copenhagen, a grande rasgos, dice: Un sistema cúantico evoluciona en una superposición de estados y cuando es medido por un objeto clásico \emph{colapsa} a uno de los estados permitidos por el objeto clásico. Volviendo al caso de la música: supongamos que yo y el minicomponente somos sistemas cuánticos. Esto es que podemos estar en superposiciones de escuchar/no-escuchar \emph{Where is my Mind} y de estar/no-estar puesto el disco correcto. Podemos estar, si nadie nos mide, realmente en ambos estados al mismo tiempo y no solamente en uno u otro. Ahora, supongamos que mi novio es un objeto clásico y me pregunta -¿Escuchaste \emph{Where is my Mind}?-. Al hacerlo me está midiendo y yo \emph{colapso} a uno de los estados clásicos que él me permite y le digo si o no. Lo mismo pasa si va a revisar el minicomponente: el ve o no ve el disco, pero no ve una superposición disco/no-disco porque el sólo permite disco o no-disco pero no superposiciones. 

Siguiendo con este ejemplo y volviendo al tema de las correlaciones. Si yo y el minicomponente estuviésemos máximamente correlacionados en el sentido clásico estaríamos con certeza absoluta en alguno de los estados ``disco y escucho'', ``disco y no-escucho'', ``no-disco y escucho'' o ``no-disco y no-escucho'', pero cúanticamente podemos estar con certeza absoluta en algún estado que sea una superposición entre, por ejemplo, ``disco y escucho'' y ``no-disco y no-escucho''. Este tipo de estados está más correlacionado que los anteriores y no son admisibles en una interpretación clásica del mundo.  A este tipo de correlaciones que van más allá de los límites de la física clásica las llamamos \emph{Entrelazamiento}.

Para que todo esto tenga sentido tiene que haber algún objeto clásico que pueda medir o reconocer este tipo de superposiciones. Continuando con el ejemplo esto podría ser un amigo especial que no pueda reconocer si estoy escuchando o no sino que reconozca si estoy en un estado escuchando/no-escuchando y que al ser medido colapse a este. El principio de incerteza aplicado a este ejemplo implica que no puede haber un aparato que mida simultaneamente superposiciones y estados ``normales''. Es decir que si mi amigo tiene como estado permitido la superposición escucho/no-escucho le son prohibidos los estados escuchando y no escuchando por separado. Esto último por más raro que parezca en el contexto de este ejemplo, en física atómica puede involucrar simplemente rotar un instrumento de medida. 

Esta intrigante propiedad de la mecánica cuántica fue señalada como una paradoja por Schr\"{o}dinger, como una razón para decir que la mecánica cuántica no era completa por Einstein y sus ``secuaces''\footnote{Einsten, Podolsky, Rosen\cite{EPR}}, y finalmente por muchos otros físicos como una de las propiedades fundamentales de la mecánica cuántica. En 1964 Bell\cite{Bell} estableció criterios matemáticos estrictos para los límites que pueden tener las correlaciones clásicas y propuso varios experimentos que se realizaron con éxito \cite{PhysRevLett.49.91}\cite{PhysRevLett.49.1804}, en los que se miden correlaciones que superan las clásicamente permitidas, demostrando que eran incorrectas las suposiciones en las que se basaban las observaciones de ``los reaccionarios'' Einstein, Podolsky y Rosen. El tema del gato de Schr\"{o}dinger fue recién totalmente comprendido cuando se desarrolló la teoría de la decoherencia\footnote{Ver apéndice \ref{ap.decoherencia}.}.

\section{Computación Cuántica}
\label{sec-comcuan}
Doy a continuación una breve descripción de algunos aspectos fundamentales de la Computación Cuántica que creo fundamentales para entender el contexto en el que fue pensado este trabajo. Para una exposición mucho más completa ver el libro \emph{Quantum Computation and Quantum Information} de Nielsen Y Chuang \cite{NC} o las \emph{Lecture Notes} de John Preskil \cite{Presk}.

\subsection{Bits y qubits y compuertas}
El mundo descripto por la mecánica cuántica admite estados que son superposición de los estados clásicamente permitidos. En este sentido es natural pensar que una computadora cuántica tendrá entonces como variables estados que son superposición de los estados permitidos de una computadora clásica, esto es superposiciones de los valores que puede tener un bit: 0 o 1. A estas nuevas variables las llamamos qubits.

Los posibles estados de un bit clásico se rotulan simplemente con los números 0 y 1. En computación cuántica para indicar como se encuentra un qubit se debe indicar en que estado está el sistema cuántico que guarda esta variable. Una de las notaciones más comunes para indicar estados cuánticos es la notación de \emph{bra-kets} de Dirac. En esta un sistema que está en estado $x$ se lo rotula $\ket{x}$. Un qubit $\ket{Q}$ que es una superposición de los estádos $\ket{0}$ y $\ket{1}$ de una computadora clásica se escribe cómo:
\begin{eqnarray}
\ket{Q}=\alpha\ket{0}+\beta\ket{1}&\mbox{donde}&|\alpha|^2+|\beta|^2=1
\end{eqnarray}
Así como para sistemas cuánticos se generalizó la noción de bit a qubit, se generaliza la idea de compuertas lógicas. 
\subsubsection{Compuertas de un qubit}
La única compuerta lógica clásica de un bit no trivial que se puede definir es la NOT, que cambia el estado del bit. En computación cuántica se la generaliza diciendo que actúa linealmente sobre el estado. Si llamamos $X$ a esta compuerta se tiene:
\begin{eqnarray}
X\ket{Q}=X(\alpha\ket{0}+\beta\ket{1})=\alpha\ket{1}+\beta\ket{0}
\end{eqnarray}
Si en vez de usar la notación de bra-kets utilizamos una matricial en la base computacional (la de los kets $\ket{0}$ y $\ket{1}$) la compuerta NOT se escribe como:
\begin{eqnarray}
X\equiv
\left[\begin{array}{cc}0&1\\1&0\end{array}\right]
&\mbox{de modo que}&
X\left[\begin{array}{c}\alpha\\\beta\end{array}\right]=\left[\begin{array}{c}\beta\\\alpha\end{array}\right]
\end{eqnarray}
Visto de esta manera es evidente que en computación cuántica podemos hacer otras compuertas de un qubit no triviales. En general como la compuerta, físicamente, es la evolución del sistema dada por un Hamiltoniano estas podrán ser todas las matrices unitarias de $2\times2$. Cualquiera de ellas que puede ser descompuesta en tres rotacines y una fase.\\
En particular una compuerta muy relevante para la computación cuántica es la Hadamard $H$.
\begin{eqnarray}
H\equiv\frac{1}{\sqrt{2}}
\left[\begin{array}{cc}1&1\\1&-1\end{array}\right]
&\mbox{de modo que}&
H\left[\begin{array}{c}\alpha\\\beta\end{array}\right]=
\frac{1}{\sqrt{2}}\left[\begin{array}{c}\alpha+\beta\\\alpha-\beta\end{array}\right]
\end{eqnarray}
o bien
\begin{eqnarray}
H(\alpha\ket{0}+\beta\ket{1})=\alpha\frac{\ket{0}+\ket{1}}{\sqrt2}+\beta\frac{\ket{0}-\ket{1}}{\sqrt2}
\end{eqnarray}

\subsubsection{Compuertas de varios qubits}
Las compuertas de más de un qubit clásicas suelen tomar una cierta cantidad de bits de entrada y dar como resultado de una operación lógica una cantidad menor de bits. Estas no son reversibles en el sentido que al hacer la operación hay información que se pierde y la entropía del sistema aumenta\footnote{Existen también lógicas clásicas reversibles pero las usuales no lo son.}. Las compuertas cuánticas son, por lo general, definidas como la extensión reversible y lineal de su correspondiente clásico.\\
En particular presento una compuerta de dos qubits relevante: la \emph{controlled not} (CNOT) es la extensión lineal de una compuerta que toma dos qubits y niega el segundo sólo si el primero esta en el estado $\ket{1}$ y no cambia el primer qubit. Su representación matricial en la base computacional es:
\begin{eqnarray}
U_{CN}\equiv
\left[\begin{array}{cccc}1&0&0&0\\0&1&0&0\\0&0&0&1\\0&0&1&0\end{array}\right]
\end{eqnarray}

\subsubsection{Compuertas Universales}
La relevancia de la CNOT viene de que, al igual que con la compuertas clásicas AND OR y NOT puede construirse cualquier compuerta clásica, con una composición finita de CNOT y compuertas de un qubit puede aproximarse tanto como se desee cualquier compuerta cuántica. La demostración de lo enunciado esta didácticamente expuesta en el libro de Nielsen y Chuang \cite{NC} donde además se señalan las referencias a los trabajos en los cuales se demostró esto originalmente.

\subsection{Algoritmos}
A groso modo hay tres cases de algoritmos eficientes en computación cuántica. Por un lado los basados en la transformada de Fourier discreta entre los cuales están el famoso algoritmo de factorización de Shor, algoritmos que encuentran periodos de funciones, estimadores de fases, etc. Por otro lado algoritmos de búsqueda como el de Grover y finalmente algoritmos de simulación de sistemas físicos complicados. Presentaré aquí la versión más simple de un algoritmo del primer tipo - el algoritmo de Deutsh - que no tiene ninguna importancia real en el sentido de la eficiencia pero es fácil de entender y muestra algunas de las ideas fundamentales\footnote{Para más detalle y otros algoritmos ver el libro de Nielsen y Chuang \cite{NC}.}. A los tipos de algoritmos mencionados deben sumarseles los, para nada menos importantes, códigos de teleportación cúantica mediante los cuales se pasa \emph{toda} la información de un cierto estado a otro separado espacialmente del primero. 

\subsubsection{Procesamiento en paralelo}
Como mencioné anteriormente el primer trabajo que propuso un algoritmo que una computadora cuántica puede resolver más eficientemente que una clásica fue el de Deutsch. Este es en realidad un algoritmo algo inútil pero igualmente significante. Describo a continuación una versión simplificada y mejorada. La idea es tratar de encontrar si una función $f:\{0,1\}\rightarrow\{0,1\}$ es constante o no. Es decir si $f(1)=f(0)$ o $f(1)\neq f(0)$. Clásicamente se requeriría ``preguntarle'' a $f$ dos veces cual es su imagen. Osea calcular $f(0)$ y luego $f(1)$. Cuánticamente veremos que con sólo ``preguntarle'' una vez basta. El algoritmo parte de un estado inicial de dos qubits.
\begin{equation} \ket{Q_0}=\ket{0}\ket{1}\label{eq.Deut.0}\end{equation}
Se le aplica una compuerta Hadamard a cada uno de los qubits.
\begin{equation}H_1\otimes H_2\ket{Q_0}=\ket{Q_1}=
\left[	\frac{\ket{0}+\ket{1}}{\sqrt2}\right]
\left[  \frac{\ket{0}-\ket{1}}{\sqrt2}\right]
\label{eq.Deut.1}
\end{equation}
Luego se aplica una compuerta $U_f$ que evalúa una función\footnote{Debe entenderse que esta función es lineal en la superposición de estados.} que al segundo qubit lo lleva a $y\oplus f(x)$ y al primero no lo modifica (donde $y$ es el valor del primer qubit y $x$ del segundo y $\oplus$ es la suma módulo 2).

\begin{eqnarray}
\label{eq.Deut.2}
U_f\ket{Q_1}=\ket{Q_2}=
\left\{\begin{array}{lc}
		\pm\left[\frac{\ket{0}+\ket{1}}{\sqrt2}\right]
		\left[\frac{\ket{0}-\ket{1}}{\sqrt2}\right]
			&\mbox{si $f(1)=f(0)$}\\\\
		\pm\left[\frac{\ket{0}-\ket{1}}{\sqrt2}\right]
		\left[\frac{\ket{0}-\ket{1}}{\sqrt2}\right]
			&\mbox{si $f(1)\neq f(0)$}
	\end{array}
\right.
\end{eqnarray}
Por último se le aplica una compuerta Hadamard al primer qubit.
\begin{eqnarray}
H_1\ket{Q_2}=\ket{Q_3}=
\left\{\begin{array}{lc}
		\pm\ket{0}
		\left[\frac{\ket{0}-\ket{1}}{\sqrt2}\right]
			&\mbox{si $f(1)=f(0)$}\\\\
		\pm\ket{1}
		\left[\frac{\ket{0}-\ket{1}}{\sqrt2}\right]
			&\mbox{si $f(1)\neq f(0)$}
	\end{array}
\right.\label{eq.Deut.3}
\end{eqnarray}
Y ya está: midiendo el primer qubit se sabe si la función es constante o no. El argumento en contra de este procedimiento como computadora cuántica eficiente es que, a pesar de que uno sólo tiene que ``preguntarle''  a $f$ cuanto vale sólo una vez, se deben usar dos qubits y aplicar Hadamards antes y después. Es cierto, este no es un circuito muy eficiente pero fue el primero e ilustra claramente la idea de procesamiento en paralelo de una computadora cuántica.\\

\section[Entrelazamiento y Computación Cuántica]{Entrelazamiento y Computación\newline Cuántica}
He dado hasta ahora una noción de lo que es el entrelazamiento y de problemas que la computación cuántica se propone resolver. El objetivo de esta sección es formalizar la idea de entrelazamiento y finalmente relacionarlo con computación cuántica. 
\subsection{Separabilidad}
Cómo primera instancia en la formalización de las ideas sobre el entrelazamiento es fundamental introducir el concepto de estado separable. Un estado es no entrelazado cuando es separable y viceversa. Para hablar de entrelazamiento y separabilidad es obviamente necesario pensar en sistemas constituidos por más de una parte -subsistemas- ya que sino no habría entre que generar entrelazamiento o que separar de que. Es decir algo no puede estar entrelazado con sigo mismo o separado de si mismo.

Un sistema de objetos distinguibles\footnote{Siempre consideraremos que los objetos que se describen son distinguibles. Esta suposición tiene sentido en el marco de la computación cuántica donde se requiere poder distinguir un qubit de otro. Por más que usualmente se piensa en usar algún grado de libertad de partículas idénticas como qubit se considera que cada una está confinada espacialmente. Las definición de estado separable para un sistema de partículas indistinguibles es distinta a la que se da aquí.} es separable si su estado, descripto por un ket $\ket{\psi}$, puede ser escrito como un producto directo de los kets que describen cada subsistema\footnote{El subsistema $i$ en el estado $m_i$ es descripto por el ket $\ket{m_i}_i$. En el caso de qubits los $m_i$ pueden valer sólo 0 o 1.}. De modo que un estado separable puede ser escrito en alguna base de la forma.
\begin{eqnarray}
\ket{\psi}&=&\ket{m_1}_1\ket{m_2}_2\ldots\ket{m_n}_n\\
&=&\ket{m_1m_2\ldots m_n}
\end{eqnarray}
Un estado es no separable o entrelazado si no puede escribirse de esta manera.
\subsubsection{Separabilidad - Matrices densidad}
La definición anterior anterior de estados separables es válida sólo para estados puros. Para describir estados mixtos es necesario introducir el formalismo de la \emph{matriz densidad}. En términos de matrices densidad un sistema de dos componentes $A$ y $B$ descripto por una matriz densidad $\rho$ es separable si\cite{PhysRevA.40.4277} 
\begin{eqnarray}\rho=\sum_\alpha q_\alpha \rho_A^\alpha\otimes\rho_B^\alpha&;&q_\alpha>0
\end{eqnarray}
Esto implica que un estado general (puro o mixto) es separable si puede ser escrito como una superposición estadística de estados puros separables. Es decir si existe alguna base en la cual puede ser escrito de esta manera.
\subsection{Estados de Bell}
Un caso muy importante de estados entrelazados son los que se llaman estados de Bell, que a veces se esconden bajo los seudónimos de pares o estados EPR\footnote{Nombre debido a los ``reaccionarios'' Einstein, Podolsky y Rosen}. Los estados de Bell son estados de un sistema compuesto por dos subsistemas -\emph{sistema bipartito}- de dimensión dos. Estos son:
\begin{subequations}\label{ec:bell}\begin{eqnarray}
\ket{\beta_{00}}=\frac{\ket{00}+\ket{11}}{\sqrt2}\\
\ket{\beta_{01}}=\frac{\ket{01}+\ket{10}}{\sqrt2}\\
\ket{\beta_{10}}=\frac{\ket{00}-\ket{11}}{\sqrt2}\\
\ket{\beta_{11}}=\frac{\ket{01}-\ket{10}}{\sqrt2}
\end{eqnarray}\end{subequations}
Los estados de Bell son de gran importancia ya que, como veremos en la sección \ref{sec-entop}, estos son máximamente entrelazados en el sentido que en un sistema bipartito de dimensión dos no hay estados que tengan mayor entrelazamiento que estos. Además son cuatro y son ortonormales de modo que son una buena base para este sistema. Por último menciono que se los usa como unidad de medida de entrelazamiento: un estado de Bell tiene un \emph{ebit} de entrelazamiento.
\subsubsection{Estados de Bell y compuertas}
Es fácil generar estados entrelazados de Bell a partir estados no entrelazados usando las compuertas descriptas en la sección \ref{sec-comcuan}. Por ejemplo si partimos del estado $\ket{11}$ y aplicamos una Hadamard al primer qubit se obtiene $(\ket{0} - \ket{1})\ket{0}\sqrt{2}$, que todavía es separable. Si a continuación se aplica una CNOT se obtiene $(\ket{01} - \ket{10})\sqrt{2}$ que es el estado entrelazado de Bell $\ket{\beta_{11}}$. De este mismo modo, cambiando los estados iniciales, se pueden generar todos los estados de Bell.

Del procedimiento anterior se ve que quien realmente ``genera'' el entrelazamiento es la compuerta CNOT. Sin embargo la aplicación de la compuerta CNOT directamente sobre el estado $\ket{11}$ lo lleva al $\ket{10}$ que tampoco es entrelazado. Volviendo a las ideas de la introducción esto puede explicarse de la siguiente manera. El estado $\ket{11}$ tiene correlaciones clásicas entre el valor de cada qubit. La compuerta CNOT sobre este cambia las correlaciones entre un qubit y el otro. Es decir, al alimentar la CNOT con estados de la base computacional esta los hace interactuar como sistemas clásicos y no genera entrelazamiento. En cambio si alimentamos a la CNOT con un estado que es superposición de estados en la base computacional esta reconoce las características cúanticas del sistema y, al hacerlos interactuar, genera entrelazamiento. El entrelazamiento se genera en la interacción de sistemas cuánticos.
\subsection{Entrelazamiento y Computación Cuántica}
El algoritmo de procesamiento paralelo presentado en la sección \ref{sec-comcuan} puede parecer un ejemplo bastante malo a los fines de mostrar cómo la computación cuántica usa el entrelazamiento como recurso. Explicaré ahora porque no es tan así y de este modo trato de dar una idea del rol que cumple el entelazamiento en los algoritmos de computación cuántica. Las bases de la observación mencionada se basa en que en todos los pasos del algoritmo los estados son separables (ver ecuaciones \ref{eq.Deut.0}, \ref{eq.Deut.1}, \ref{eq.Deut.2}, \ref{eq.Deut.3}). Sin embargo examinando la forma que tiene $U_f$ por ejemplo para el caso en que $f(1)=1$ y $f(0)=0$ se encuentra que es exactamente una compuerta CNOT y que en los otros casos no es más que composiciones CNOTs con Hadamards. Más aún los estados a los que se les aplica $U_f$ no le son propios de modo que $U_f$ los hace interactuar como objetos cuánticos y no clásicos. A pesar de que no se genera entrelazamiento en ninguna etapa del proceso la interacción entre los qubits es de origen cuántico y es esta la que permite realizar la tarea del modo que se hace. Es más, es posible que este algoritmo no sea realmente eficiente debido a no hay estados entrelazados de por medio.

Finalmente hago énfasis en que cualquier algoritmo puede descomponerse en CNOTs y compuertas de un qubit y, en particular, prácticamente todos los algoritmos eficientes de computación cuántica usan los CNOTs para generar estados entrelazados. El rol que cumple el entrelazamiento en la computación cuántica ha sido muy discutido en particular en torno a las realizaciones que se basan en técnicas de NMR -resonancia magnética nuclear-\cite{NMRtrouble}. Contrariamente el rol del entrelazamiento en los códigos de teleportación es fundamental.

\chapter{Medidas de entrelazamiento}
\noindent\textbf{¿Que se espera de una medida de entrelazamiento?}\\
Una medida de correlaciones puramente cuánticas. Es decir que distinga ente correlaciones cuánticas y clásicas y sólo considere las de origen cuántico. Una medida que distinga entre estados separables y no separables. En este camino se han dado grandes avances en los últimos 10 años. No se ha llegado a encontrar una medida universal de entrelazamiento aún. Por ejemplo Peres\cite{PhysRevLett.77.1413} encontró un criterio de separabilidad que es necesario pero no suficiente. Se han propuesto varias medidas aditivas y no aditivas\cite{GG2001}. En este trabajo me concentraré en dos medidas aditivas: la  entropía de non Neumann que sirve sólo para estados puros, y el entrelazamiento de formación que sirve para estados mixtos\cite{BDSW1996} y coincide con la anterior para estados puros. A pesar de que el entrelazamiento de formación es muy difícil de calcular para un caso general Hill y Wooters \cite{HW1997} encontraron una forma explícita de calcularla para un par de qubits como función de una cantidad que llamaron Concurrencia. En pos de unificar medidas simples de entrelazamiento damos aquí un paso más proponiendo una medida equivalente a la entropía de un qubit que sea comparable a la Concurrencia.

\section{Entropía - Estados Puros}
\label{sec-entop}
La entropía de von Neumann de un subsistema es una buena medida de entrelazamiento para estados puros. Mide el entrelazamiento entre dos subsistemas sin importar su dimensión. Daremos a continuación la definición, luego justificaremos porque ésta es adecuada y finalmente estudiaremos algunos ejemplos. 
\subsection{Definiciones}
La entropía de von Neumann adimensional\footnote{En el contexto de Teoría de la Información Cuántica se la suele definir adimensional y en base 2 para que, como medida de desorden, valga 1 para un qubit completamente desordenado y 0 para uno completamente ordenado.} de un sistema cuántico caracterizado por una matriz densidad $\rho$ se define como:
\begin{equation}S(\rho)=-Tr[\rho \log_2{\rho}] \end{equation}
La matriz densidad reducida de un subsistema se define como traza parcial sobre todas las variables que no pertenecen al subsistema.
\begin{eqnarray}\rho_{sub}=Tr_{\notin sub}[\rho]\end{eqnarray}
Supongamos un sistema bipartito, que puede ser dividido en dos subsistemas disjuntos $A$ y $B$. Toda la información del estado de cada subsistema está en las matrices densidad reducidas ($\rho_A$ y $\rho_B$) de cada subsistema.
\begin{eqnarray}\rho_A=Tr_{B}[\rho_{AB}]&;&\rho_B=Tr_{A}[\rho_{AB}] \end{eqnarray}
La entropía de un subsistema $A$ se define análogamente a la entropía para el sistema entero pero con la matriz reducida $\rho_A$  en vez de la matriz densidad entera $\rho$.
\begin{eqnarray}S(\rho_A)=-Tr[\rho_A \log_2{\rho_A}]&;&S(\rho_B)=-Tr[\rho_B \log_2{\rho_B}]  \end{eqnarray}
Además es inmediato ver, usando las propiedades de la descomposición de Schmidt (el apéndice \ref{sec-schmidt}) que los autovalores no nulos de $\rho_A$ y $\rho_B$ son iguales de modo que sus entropías son iguales $S(\rho_A)=S(\rho_B)$.
\\
Se puede definir una buena medida de entrelazamiento $E(\rho_{AB})$ entre dos subsistemas $A$ y $B$, que forman un estado puro, como la entropía de cualquiera de los subsistemas. 
\begin{eqnarray}E(\rho_{AB})=S(\rho_A)=S(\rho_B) \end{eqnarray}
\subsection{Propiedades/Justificación}
\label{sec-entop.prop}
Doy a continuación una serie de propiedades de la entropía de un subsistema, que justifican su elección como medida de entrelazamiento entre subsistemas de un estado puro. Algunas de estas las ideas salen de un resumen hecho por Bennett et. al. \cite{BDSW1996}.
\begin{description}
\item[El entrelazamiento de $A$ con $B$ es igual al de $B$ con $A$ .] Esto se\newline desprende directamente de la definición que considera un sistema bipartito puro y de la observación hecha anteriormente sobre los autovalores de los subsistemas. Osea $S(\rho_A)=S(\rho_B)$.
\item[Para estados separables $E(\rho_{AB})=0$ .]Si un estado puro es separable $\rho_{AB}=\rho_A\otimes\rho_B$ ,donde $\rho_A$ y $\rho_B$ son estados puros, de modo que $E(\rho_{AB})=S(\rho_{A})=S(\rho_{B})=0$ dado que la entropía de un estado puro es cero (ver apéndice \ref{sec-schmidt}).
\item[Los estados de Bell están máximamente entrelazados.] Un cál\-culo directo conduce a $E(\ket{\beta_{ij}}\bra{\beta_{ij}})=1$, $\forall i,j$ 
\item[Es aditiva $E(\rho_1\otimes\rho_2\otimes\ldots\rho_n)=E(\rho_1)+E(\rho_2)+\ldots E(\rho_n)$.] El entrelazamiento de sistemas un conjunto de sistemas independientes es la suma del entrelazamiento de cada sistema.\footnote{Algunos autores\cite{GG2001} argumentan que no es necesario exigir aditividad para una medida de entrelazamiento.}
\item[$E(\rho_{AB})$ se conserva frente a operaciones locales unitarias].\linebreak Transformaciones locales del tipo $U=U_A\otimes U_B$ no cambian el entrelazamiento entre los subsistemas \cite{BBPS1996}.
\item[$E(\rho_{AB})$ no se incrementa con operaciones locales no \linebreak unitarias.] Transformaciones locales no unitarias, como mediciones, sólo pueden disminuir el entrelazamiento pero no incrementarlo. 
\item[El entrelazamiento puede diluirse o concentrarse]\hspace{-0.15cm}\footnote{Esta propiedad es la base teórica utilizada para construir otras medidas de entrelazamiento como entropía de formación y la herramienta fundamental de diversos códigos corrección de errores.}\hspace{0.15cm} Esto es:\linebreak realizando solamente oparaciones locales en cada subsistema y permitiendo comunicación clásica entre ellos -LOCC\footnote{LOCC - Por sus siglas en inglés \emph{Local Operations and Classical Communication}}- se puede, a partir de $n$ copias de $\rho_{AB}$, preparar $m$ sistemas idénticos $\rho'_{AB}$ con eficiencia  $m/n \approx E(\rho_{AB})/E(\rho'_{AB})$ en el sentido asintótico de $n$ grande.\cite{BBPS1996} 
\end{description} 
\subsection{Entropía de un qubit}
La entropía de un qubit es una medida del entrelazamiento de ese qubit con el resto del sistema, por eso la llamaré entrelazamiento de uno con el resto.\\
La matriz reducida de un qubit siempre puede escribirse en la base computacional como:
\begin{equation}
\rho_1=\left[\begin{array}{cc}a&b\\b^*&1-a\end{array}\right]
\end{equation}
y tiene por autovalores $\lambda_\pm=\frac12\left(1\pm\sqrt{(1-2a)^2+|2b|^2}\right)$ de modo que la entropía de un subsistema arbitrario de un qubit es
\begin{eqnarray}
S(\rho_1)
&=&-\lambda_+\log{\lambda_+}-\lambda_-\log{\lambda_-}\nonumber\\
&=&-\lambda_+\log{\lambda_+}-(1-\lambda_+)\log{(1-\lambda_+)}
\end{eqnarray}
En el caso que el sistema completo tenga paridad de espín definida (ver apéndice \ref{chap-par}) se tiene que $b=0$ y la entropía sólo depende sólo de $a$ y es simplemente
\begin{equation}
S(\rho_1)=-a\log a-(1-a)\log(1-a)
\end{equation}
Además $a= \braketo{S_z}+1/2$  de modo que si $\rho_1$ conmuta con P alcanza con conocer $\braketo{S_z}$ para obtener una medida de la entropía.

\section{Entrelazamiento de Formación - Estados Mixtos}
El entrelazamiento de Formación es una buena medida de entrelazamiento para estados mixtos\cite{BDSW1996}. Mide el entrelazamiento entre dos subsistemas sin importar su dimensión. Tiene la inconveniencia de ser muy difícil de calcular. Es una extensión de la entropía, como medida de entrelazamiento, para estados mixtos.\\
La idea detrás de la definición se basa en los mecanismos de destilación y dilución mencionados en la sección anterior y la idea egipcia elegir una unidad estandar de referencia, así como el kilogramo o el metro patrón. Se toma como unidad estándar de referencia de entrelazamiento el entrelazamiento de un singlete puro. Por mecanismos de destilación se pueden generar $n$ singletes puros a partir de $m$ copias de un estado mixto o bien diluyendo se generan $m$ estados mixtos a partir de $n$ singletes puros. El entrelazamiento de formación de un estado es el cociente $m/n$ en el sentido asintótico de $n$ grande.\\

\subsection{Definición}
Para un sistema bipartito puro el entrelazamiento de formación  $E(\rho_{AB})$ es la entropía de von Neumann de uno de los subsistemas.
\begin{eqnarray}E(\rho_{AB})=S(\rho_A)=S(\rho_B)\end{eqnarray}
Para un ensamble de sistemas bipartitos puros $\mathcal{E}=\{p_i,\rho_{AB}^i\}$ el entrelazamiento de formación $E(\mathcal{E})$ se define como el promedio de los entrelazamientos de formación de cada estado $\rho_{AB}^i$ pesadas con la probabilidad  $p_i$ del estado.
\begin{eqnarray}E(\mathcal{E})=\sum_i p_iE(\rho_{AB}^i)\end{eqnarray}
Para un sistema bipartito mixto $\rho_{AB}$ el entrelazamiento de formación $E(\rho_{AB})$ se define como el mínimo de los $E(\mathcal{E})$ sobre todos los posibles ensambles $\mathcal{E}=\{p_i,\rho_{AB}^i\}$ que representen al estado mixto $\rho_{AB}=\sum_i p_i \rho^i_{AB}$.
\begin{eqnarray}E(\rho_{AB})=\min_\mathcal{E} \sum_i p_iE(\rho_{AB}^i)\end{eqnarray}
\subsection{Propiedades/Justificación}
En el trabajo de Bennett et. al.\cite{BDSW1996} en el que proponen al entrelazamiento de formación como medida de entrelazamiento demuestran que para generar un estado bipartito $\rho_{AB}$ sólo a partir de operaciones locales y comunicación clásica entre los subsistemas estos deben compartir el equivalente a $E(\rho_{AB})$ singletes puros. El entrelazamiento de formación $E(\rho_{AB})$ es la cantidad de entrelazamiento necesaria para \emph{formar} $\rho_{AB}$. Además demuestran que $E(\rho_{AB})$ no puede incrementarse con ningún tipo de operaciones locales y comunicación clásica. Es decir el entrelazamiento de formación tiene las mismas propiedades (ver sección \ref{sec-entop.prop}) que la entropía pero para estados mixtos.

\section{Concurrencia}
\label{sec-concurrencia}
La concurrencia es una medida de entrelazamiento derivada a partir del Entrelazamiento de Formación entre un par de subsistemas de dimensión 2 cada uno. Tiene la bondad de que es fácil de calcular. Fue originalmente propuesta por S. Hill y W.K. Wootters \cite{HW1997} y refinada por W.K. Wooters  \cite{Woo1998}.
\subsection{Definición}
La concurrencia $C$ de un sistema arbitrario de dos qubits $i,j$ se define como:
\begin{eqnarray}
C=Max[2\lambda_m - Tr R, 0]&;&R=[\rho_2^{1/2} \tilde{\rho} \rho_2^{1/2}]^{1/2}
\end{eqnarray}
Donde  $\rho_2=Tr_{\notin\{i,j\}}[\rho]$ (traza parcial sobre todas las variables que no son $i,j$); $\lambda_m$ es la máximo autovalor de R; y $\tilde{\rho_2}=4S_i^{y}S_j^y\rho_2^*S_i^{y}S_j^y$ (matriz densidad $\rho_2$ con los espines invertidos). \\
A los efectos del cálculo de la concurrencia es equivalente usar la raíz de los autovalores de la matriz \emph{no} hermítica $\tilde{\rho} \rho_2$ ya que las cuentas son más simples y los resultados los mismos. En este sentido, si $\lambda_1,\lambda_2,\lambda_3,\lambda_4$ son las raíces de estos autovalores en orden decreciente la concurrencia es simplemente\footnote{Esta ecuación también vale para los autovalores de $R$.}
\begin{equation}C=Max[\lambda_1-\lambda_2-\lambda_3-\lambda_4, 0]\end{equation}
\subsection{Propiedades/Justificación}
En los trabajos \cite{HW1997},\cite{Woo1998} Hill y Wootters  demuestran que la entropía de formación para un estado arbitrario de 2 qubits es una función monótonamente creciente de la concurrencia y por ende proponen usarla como medida de entrelazamiento. Explicitamente el resultado que obtienen es que la entropía de formación es
\begin{eqnarray}
E(M)&=&h\left(\frac{1+\sqrt{1-C^2}}{2}\right)\\
h(a)&=&-a\log a-(1-a)\log(1-a)\nonumber
\end{eqnarray}

\subsection{Cadena cíclica con paridad de espín}
Más adelante estudiaremos una cadena cíclica cuyo estado tiene paridad definida. En este caso la matriz densidad de dos qubits en la base computacional se escribe\footnote{Que haya términos iguales en la diagonal de la matriz es debido a la simetría cíclica de la cadena, los términos nulos son debido a la simetría de paridad (ver apéndice \ref{chap-par}).}:
\begin{equation}
\rho_2=\left(\begin{array}{cccc}
a&0&0&d^*\\
0&b&e&0\\
0&e&b&0\\
d&0&0&c\\
\end{array}\right)
\end{equation}
Un cálculo directo conduce al resultado:
\begin{eqnarray}\label{eq:concu.paridad}
C=2\;Max[|e|-\sqrt{ac},|d|-b,0]
\end{eqnarray}
Donde sólo uno de los valores puede ser positivo ya que la positividad de $\rho_2$ implica que $|e|\leq$ b y $|d|\leq\sqrt{ac}$.\\
Es importante también tener en cuenta como reconstruir $\rho_2$ a partir de los valores medios de los operadores relevantes.  En este sentido se calculan directamente como $\langle\hat{O}\rangle=Tr[\hat{O}\rho_2]$ los valores medios.
\begin{eqnarray}
\braketo{S_i^z}=\frac12(a-c)&;&
\braketo{S_i^zS_j^z}=\frac14(a-2b+c)\nonumber\\
\braketo{S_i^+S_j^-}=e&;&
\braketo{S_i^+S_j^+}=d \label{ec.valores_medios_concu}
\end{eqnarray}
La ecuación que falta sale de la condición $Tr[\rho_2]=1$; de este modo $1=a+2b+c$. Estas son todas las herramientas necesarias para reconstruir $\rho_2$ y de este modo la concurrencia de pares.

\begin{figure}[t]
\begin{center}
\epsfig{file=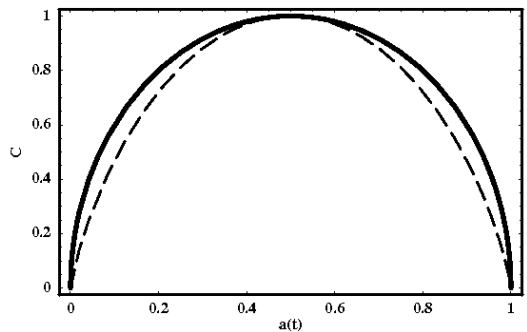, angle=0, width=200 pt}
\caption{\small Concurrencia $2\sqrt{a(1-a)}$ (línea sólida) y entrelazamiento de formación $h(a)=-a\log a-(1-a)\log(1-a)$ (línea punteada) como función de $a$ \label{fig.entropalg}}
\end{center}
\end{figure}

\subsection{Concurrencia de estados de Bell}
Los estados de Bell tienen todos máxima concurrencia $C=1$. Para los estados mixtos que conmutan con la paridad de espín(ver apéndice \ref{chap-par}) vemos que si el término de \ref{eq:concu.paridad} que da a lugar a la concurrencia es $|d|-b$ entonces el entrelazamiento corresponde a un estado par mientras que si el término relevante es $|e|-\sqrt{ac}$ corresponde a uno impar. La paridad de espín resulta muy útil para clasificar los estados y el tipo de entrelazamiento de pares en los sistemas a estudiar.

\subsection{Concurrencia de uno con el Resto}
\label{sec-entrlaza.algebra}
Tanto  el entrelazamiento de formación de un par como el de uno con el resto se escriben como:
\begin{equation}
h(a)=-a\log a-(1-a)\log(1-a)\label{ec:malditoslogs}\end{equation}
donde para el entrelazamiento de un par $a=(1+\sqrt{1-C^2})/2$ y para el entrelazamiento de uno con el resto $a=\lambda_+$ o $a=\lambda_-$. El entrelazamiento de formación de un par se reduce a la entropía de un qubit en el  caso en que el par es un estado puro. En este sentido ambas medidas son comparables pero tienen la desventaja de que haya logaritmos de por medio.\\
Debido a que $C=2\sqrt{a(1-a)}$ y que como fue señalado anteriormente $C$ y $h(a)$ son funciones monótonamente crecientes una de la otra puede definirse la concurrencia de uno con el resto de este modo pero usando $a=\lambda_+$. \\
\begin{eqnarray}
C_1(M)&=&2\sqrt{a(1-a)}=2\sqrt{\lambda_+(1-\lambda_+)}\\
C_2(M)&=&2\sqrt{a(1-a)}=C
\end{eqnarray}

Donde hemos acuñado los símbolos $C_1$ y $C_2$ para referirnos a las medidas de entrelazamiento de uno con el resto y de pares respectivamente. En lo que continúa usaré los términos entrelazamiento y concurrencia indiscriminadamente para referirme siempre a estas medidas.

Un último comentario que merece la atención es la forma de la función $2\sqrt{a(1-a)}$. En la figura \ref{fig.entropalg} se muestra esta función como la alternativa a $h(a)$. Ambas están definidas en el intervalo $0\le a\le1$ y tienen un máximo en $a=1/2$ donde valen 1.


\part{Entrelazamiento en una Cadena XY}
\chapter[Computación Cuántica en una cadena XY]{Computación cuántica en una cadena XY}
\label{chap-sist}
De los muchos posibles sistemas que se podrían elegir para hacer computación cuántica, y sobre los cuales se esta investigando actualmente \cite{roadmap}\cite{NC} , algunos pueden ser descriptos aproximadamente con Hamiltonianos XY. En particular estos podrían ser de NMR(resonancia magnletica nuclear), de Estado Sólido (basados en espín cómo en ``Quantum dots'') o superconductores. En este sentido se ha estudiado el entrelazamiento de estados térmicos para este tipo de cadenas \cite{rossignoli:012335}. 
También se han propuesto esquemas mediante los cuales se podrían implementar compuertas cuánticas en cadenas de este tipo de sistemas\cite{benjamin:247901}.
\subsection{El Hamiltoniano}
El sistema que estudié es una cadena cíclica de $n$ espines interactuando por medio de acoplamientos de Heisenberg del tipo XY a primeros vecinos en presencia de un campo magnético uniforme y constante perpendicular a la dirección de las interacciones. El Hamiltoniano para este sistema es
\begin{equation}
H=\sum_{i=1}^n \bigl[ b S_i^z - v_x S_i^xS_{i+1}^x- v_y S_i^yS_{i+1}^y\bigr]
\end{equation}
donde $n+1\equiv1$.
Definiendo los operadores usuales de subida $S^+$ y de bajada $S^-$ como:
\begin{eqnarray}
S^+_i=S^x_i+iS^y_i &;& S^-_i=S^x_i-iS^y_i
\end{eqnarray}
el Hamiltoniano queda expresado como:
\begin{eqnarray}
H&=&\sum_{i=1}^n \bigl[  b S_i^z - v(S_i^+S_{i+1}^-+S_{i+1}^+S_{i}^-)  -g(S_i^+S_{i+1}^+ +S_{i+1}^-S_{i}^-)\bigr] \nonumber \\
&=&\sum_{i=1}^n \bigl[  b S_i^z - (vS_i^+S_{i+1}^- +gS_i^+S_{i+1}^+ +h.c.)\bigr] 
\end{eqnarray}
Donde $(v,g)=(v_x\pm v_y)/4$. A $v$ lo llamaré parámetro de ``hopping'' y a $g$ anisotropía.
\subsection{La condición inicial}
En todos los casos estudiaré la evolución temporal del entrelazamiento de un sistema que inicialmente tiene todos los espines alineados antiparalelos al campo magnético. Tiene sentido estudiar esta condición inicial ya que es experimentalmente muy fácil de generar. Un campo magnético suficientemente fuerte siempre va a alinear los espines en una misma dirección. La función de onda a $t=0$ es 
\begin{eqnarray}
\ket{\Psi(0)}=\ket{\sd\sd\ldots\sd}\label{ec:psi.inicial}
\end{eqnarray}
el estado del sistema como función de $t$ es (se asume en lo sucesivo $\hbar=1$)
\begin{eqnarray}
\ket{\Psi(t)}=e^{-iHt}\ket{\sd\sd\ldots\sd}\label{ec:psi.t}
\end{eqnarray}

\subsection{Paridad de espín}
\label{sec-parham}
\subsubsection{Paridad y Hamiltoniano}
Es simple ver que Hamiltoniano conmuta con el operador paridad definido en el apéndice \ref{sec-parspines}. 
\begin{eqnarray}
P=e^{i\pi(\sum_i S^z_i + n/2)}
\end{eqnarray}
Recordemos que este operador distingue la paridad de un estado si hay una cantidad par o impar de espines para ``arriba''.
El Hamiltoniano tiene términos $S^z_i$ es evidente que  conmutan con $P$. Tiene también términos tipo $S_i^+S_{i+1}^-$ y $S_i^+S_{i+1}^+$. Los primeros tampoco cambian la paridad ya que ``suben'' un espín y ``bajan'' otro sin alterar la cantidad de espines que hay ``arriba''. Osea, sin modificar la paridad. Los segundos ``suben'' dos espines por lo que tampoco cambia la paridad. Como ningún término del Hamiltoniano cambia la paridad de una función de onda dada se tiene que:
\begin{eqnarray}
[P,H]=0
\end{eqnarray}
\subsubsection{Paridad y la evolución temporal}
El estado inicial conmuta con $P$ (tiene paridad definida). Esto se ve directamente de su definición \ref{ec:psi.inicial}.
Como $P$ conmuta con $H$ y con $\rho(0)$ se obtiene inmediatamente que $\rho(t)$ tiene paridad definida para cualquier $t$.

\section[Cadenas XY Cortas]{Diagonalización directa: \newline Entrelazamiento en cadenas cortas}
\subsection{Dos qubits}
\label{sec-dosqbits}
\begin{figure}[h]
\label{fig:dosqbits}
\begin{center}
\epsfig{file=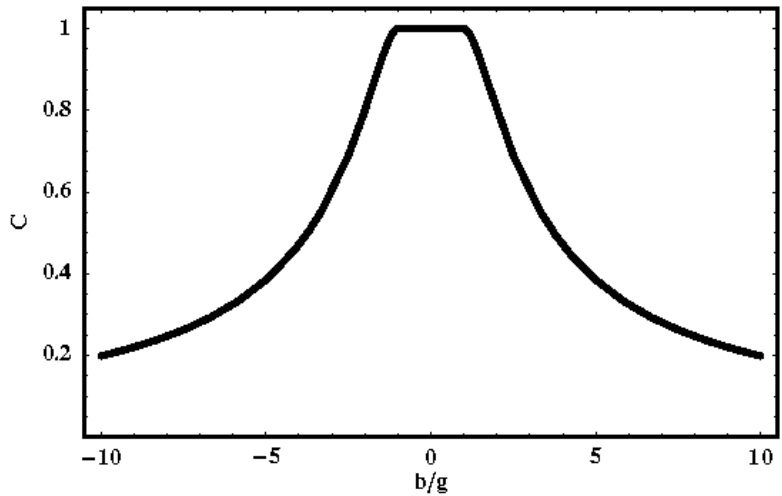, angle=0, width=300 pt}
\caption{\small Entrelazamiento $C_{1,2}$ máximo como función de $b/g$ para dos qubits.}
\end{center}
\end{figure}

Empezaremos estudiando el caso simple de 2 qubits. Por las razones mencionadas en la sección \ref{sec-parham} el sistema evolucionará dentro del subespacio de paridad definida que contiene al estado inicial $\ket{\sd\sd}$. Este es el subespacio de paridad positiva. En la base normal de autovectores de $S_z$ ($\ket{\su\su}$ y $\ket{\sd\sd}$) el Hamiltoniano en forma matricial es:
\begin{equation}
H=\left(\begin{array}{cc}b&-g\\-g&-b\end{array}\right)
\end{equation}
que tiene autovalores $E_\pm=\pm\lambda$ con
\begin{eqnarray}
\lambda=\sqrt{b^2+g^2}
\end{eqnarray}
de modo que la evolución temporal del sistema, dada por la ecuación \ref{ec:psi.t}, es 
\begin{eqnarray}
\ket{\Psi(t)}=i\frac{g}{\lambda}\sin(\lambda t)\ket{\su\su}
+\left[\cos(\lambda t)+i\frac{b}{\lambda}\sin(\lambda t)\right]\ket{\sd\sd}\label{eq:dosqbits.ev.temp}
\end{eqnarray}
En este caso al haber sólo dos qubits el entrelazamiento de pares es el mismo que el de uno con el resto: $C_1=C_2$. Para calcularlo es necesario conocer la función $a(t)$ (ver sección \ref{sec-entrlaza.algebra}). Esta es simplemente una lorenziana modulada con centro en $b=0$.
\begin{eqnarray}
a(t)=\frac{g^2}{b^2+g^2}\sin^2(\lambda t)
\end{eqnarray}
y su máximo se da para $t=\frac{(2m+1)\pi}{2\lambda}$, $m\in\mathcal{R}$
\begin{eqnarray}
a_m=\frac{g^2}{b^2+g^2}
\end{eqnarray}
El estado máximamente entrelazado de dos qubits (se da cuando $a(t)=1/2$. Si $a_m>1/2$ siempre se logra un estado de Bell para $t_m=\lambda^{-1}\arcsin((2a_m)^{-1/2})$; este caso se da cuando  $|b|/g<1$. Si $a_m<1/2$ nunca se llega a generar un estado de Bell perfecto (sin ruido), el entrelazamiento máximo se da para $t_m=\pi/2\lambda$ cuando $a(t)=a_m$.
La figura \ref{fig:dosqbits} muestra el entrelazamiento máximo $2\sqrt{a(1-a)}$ alcanzado en función de $b/g$. Se observa una meseta estricta -\emph{plateau}- para $|b|/g<1$ y un comportamiento asintótico para $|b|/g\gg 1$ de la forma $2g/|b|$. El entelazamiento en este caso no depende del parámetro de ``hopping'' $v$ sino que  sólo depende de la relación entre la anisotropía de la interacción $g$ y el campo magnético $b$.

\begin{figure}[h]
  \begin{center}
    \mbox{
      \subfigure[$b=0.2$]{\scalebox{0.5}{\epsfig{file=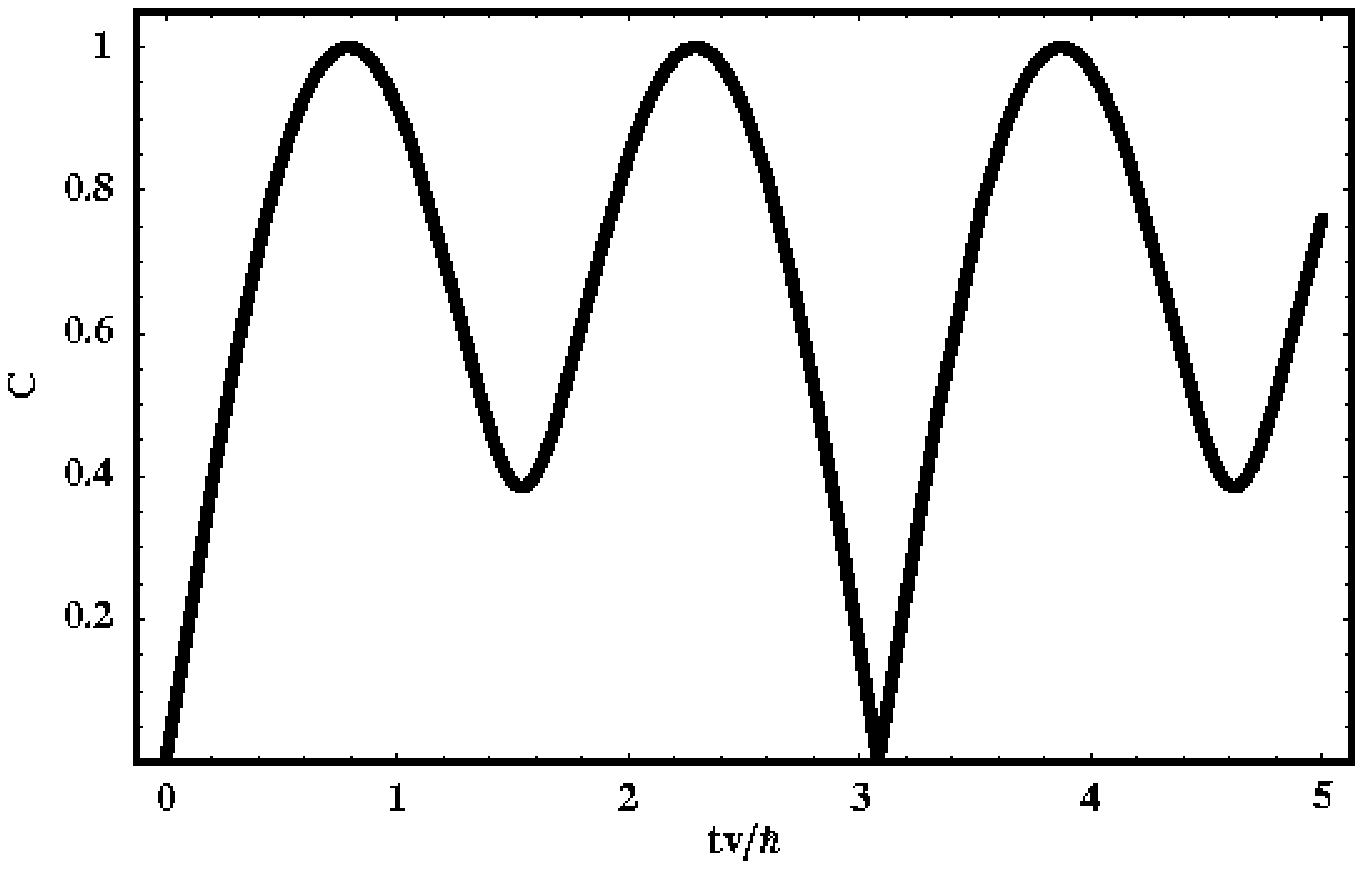}}\label{fig:tdosqbits02} }\quad
      \subfigure[$b=1$]{\scalebox{0.5}{\epsfig{file=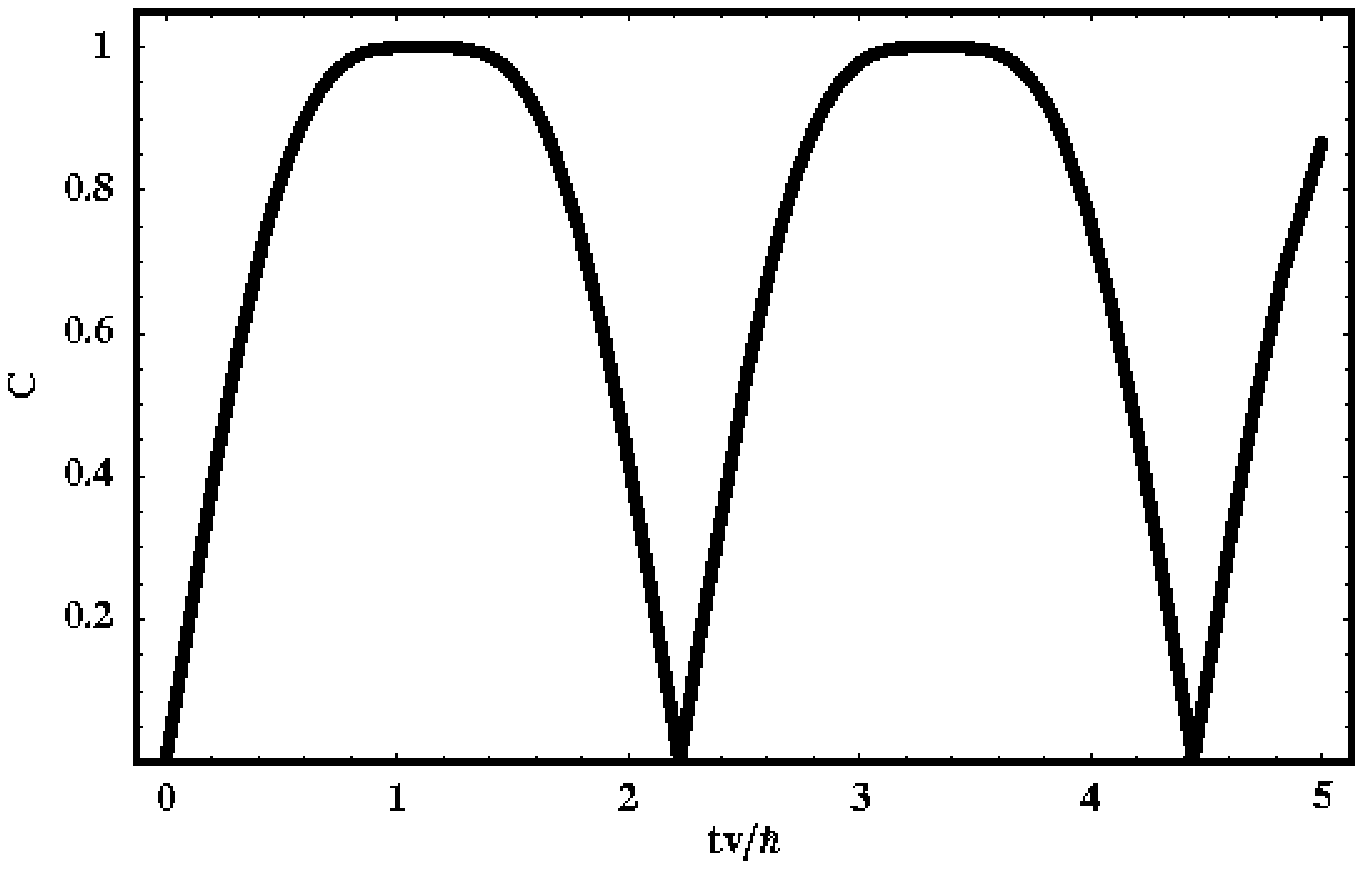}}\label{fig:tdosqbits10}}
      }
    \mbox{
      \subfigure[$b=2$]{\scalebox{0.5}{\epsfig{file=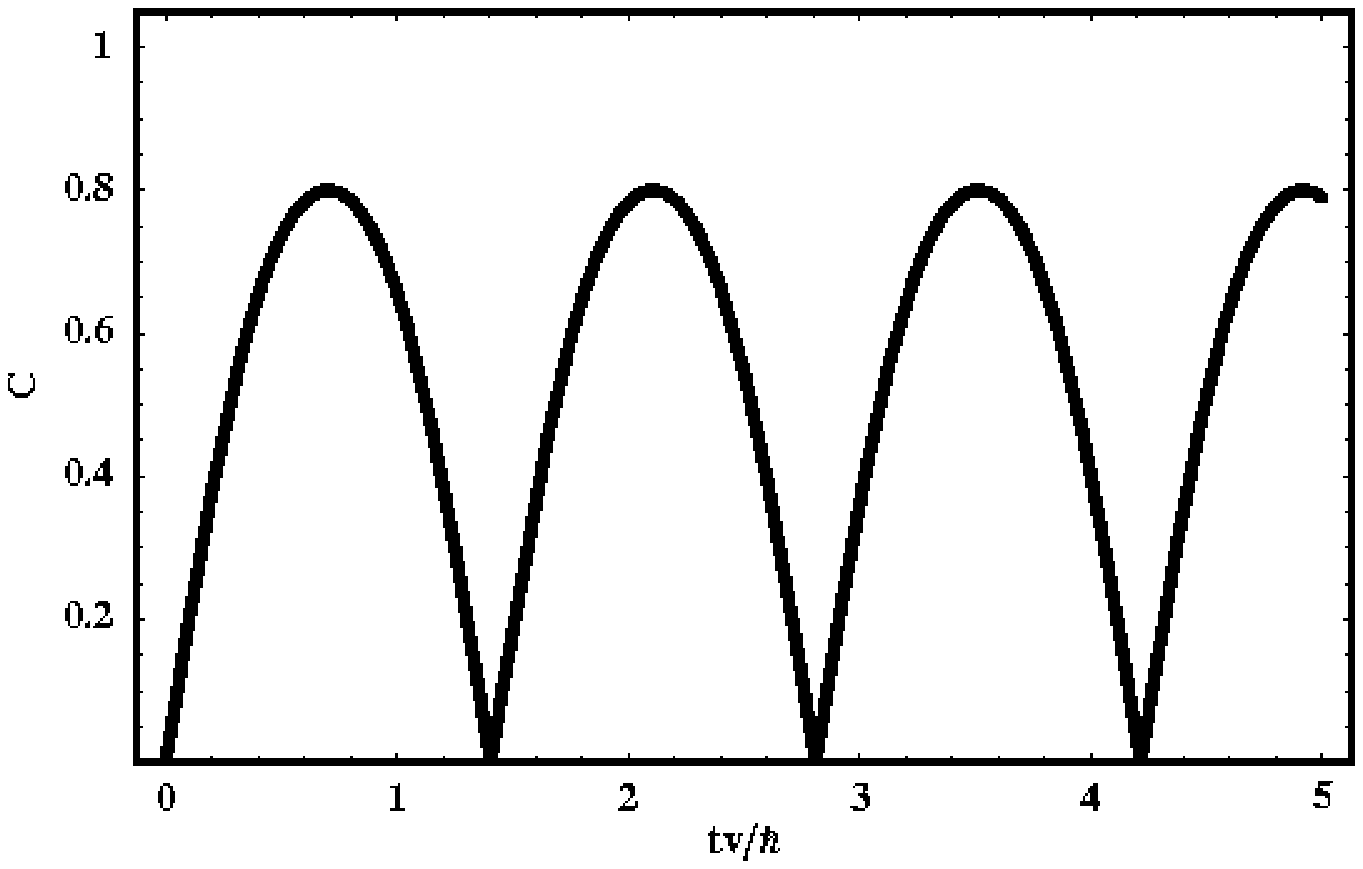}}} \quad
      \subfigure[$b=4$]{\scalebox{0.5}{\epsfig{file=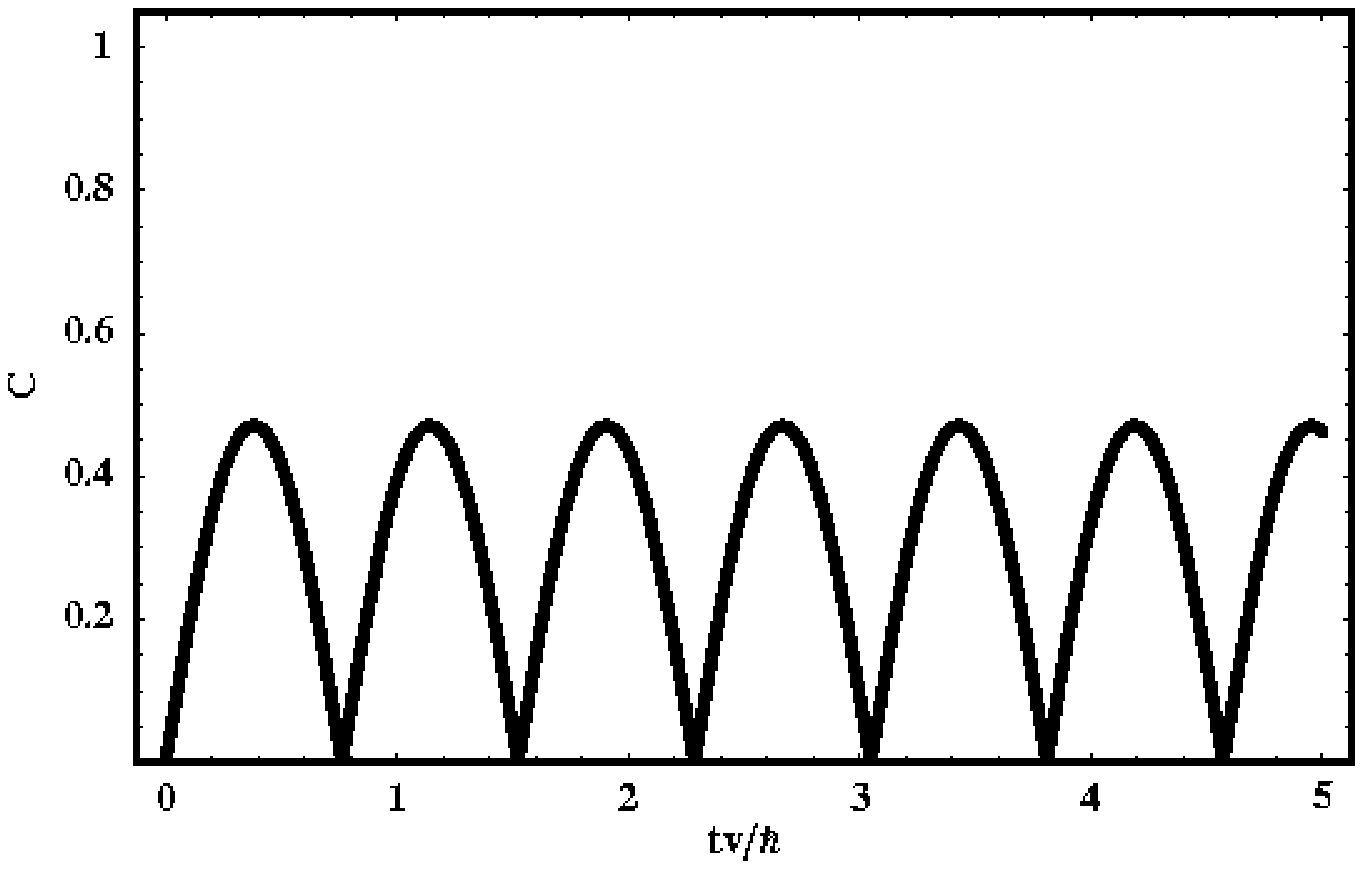}}} 
      }
    \caption{\small Entrelazamiento $C_{1,2}$ en función del tiempo en unidades de $\hbar/v$ para dos qubits y distintos valores de $b/g$.}
    \label{fig:tdosqbits}
  \end{center}
\end{figure}

Veamos las características
más importantes de la evolución temporal para distintos rangos de $b/g$. De la ecuación \ref{eq:dosqbits.ev.temp}, que rige la evolución temporal de este sistema y de la  condición $a(t)=1/2$ se ve que los estados máximamente entrelazados que se pueden obtener son
\begin{eqnarray}
\pm\frac{i}{\sqrt2}\left(\ket{\su\su}+e^{\pm i\phi}\ket{\sd\sd}\right)& ; &\cos{\phi}=\frac{b}{g}
\end{eqnarray}
Este estado no es ninguno de los estados de Bell en la base normal (\ref{ec:bell}) pero tiene el mismo entrelazamiento que un estado normal de Bell y tiene paridad positiva, no es más que un estado de Bell rotado y con una fase. Además se ve claramente, y puede verificarse con las figuras \ref{fig:tdosqbits}, que para $b/g\gg1$ el término relevante es el que tiene al estado inicial $\ket{\sd\sd}$ y por lo tanto el entrelazamiento es pequeño; mientras que para $b/g<1$ los términos relevantes son los que tienen los senos, que multiplican tanto a $\ket{\sd\sd}$ como a $\ket{\su\su}$, de modo que efectivamente se genera un estado de Bell.\\ 

Una particularidad interesante para notar, en vistas de aplicar esto a sistemas reales que funcionen como computadoras cuánticas, es que el tiempo que permanece el sistema con $C_{1,2}\approx1$ es mucho mayor cuando $b/g=1$ (fig \ref{fig:tdosqbits10}). Es decir que no por agrandar la anisotropía se tienen mejores condiciones de formación de entrelazamiento ya que para $b/g<1$  el tiempo que permanece el sistema en $C_{1,2}\approx1$ es menor (fig \ref{fig:tdosqbits02}).  En sistemas de ``Quantum dots'', por ejemplo, se han reportado\cite{PhysRevLett.83.4204} valores de $g\approx0.1$meV que implican una escala de tiempo del orden de las decenas de pico segundos ($10^-11s$).

\subsection{Tres qubits}
\label{sec.tres_qbits}
Seguimos estudiando el caso un poco más complejo de 3 qubits. En este caso el estado inicial es $\ket{\sd\sd\sd}$ por lo que el subespacio relevante es el generado por los vectores $\ket{\sd\sd\sd}$ ,$\ket{\sd\su\su}$,$\ket{\su\su\sd}$ y $\ket{\su\sd\su}$. De los últimos tres estados, debido a la simetría cíclica del Hamiltoniano, sólo se puebla la combinación simétrica de ellos y por lo tanto el espacio relevante tiene, todavía, dimensión 2. Elegimos para este subespacio la base que contiene al estado inicial $\ket{\sd\sd\sd}$ y al estado de Werner\cite{PhysRevA.40.4277} $\ket{\mathcal{W}}=\left(\ket{\sd\su\su}+\ket{\su\su\sd}+\ket{\su\sd\su}\right)/\sqrt3$.  El Hamiltoniano es en esta base 
\begin{equation}
H=\left(\begin{array}{cc} \frac12b-2v&-\sqrt3g\\-\sqrt3g&-\frac32 b\end{array}\right)
\end{equation}
que tiene autovalores $E_\pm=E_0\pm\lambda$ con
\begin{eqnarray}
E_0=-\frac{b}2-v&;&\lambda=\sqrt{(b-v)^2+3g^2}
\end{eqnarray}
de modo que la evolución temporal del sistema, que viene dada por la ecuación \ref{ec:psi.t}, es 
\begin{eqnarray}
\ket{\Psi(t)}=e^{iE_0t}\left[\cos(\lambda t)+i\frac{b-v}{\lambda}\sin(\lambda t)\right]\ket{\su\su\su}
+\frac{\sqrt3g}{\lambda}\sin(\lambda t)\ket{\mathcal{W}}
\end{eqnarray}
En este caso, al haber tres qubits, hay diferencia entre el entrelazamiento de uno con el resto y de pares, sin embargo veremos que están íntimamente relacionados y que los resultados que se obtienen pueden entenderse muy bien en términos intuitivos de entrelazamiento. La función $a(t)$ es, como antes, una lorenziana modulada pero con centro en $b=v$, ancho $g$ y valor máximo $2/3$.
\begin{eqnarray}
a(t)=\frac{2g^2}{(b-v)^2+3g^2}\sin^2(\lambda t)
\end{eqnarray}
y su máximo se da para $t=\frac{(2m+1)\pi}{2\lambda}$, $m\in\mathcal{R}$
\begin{eqnarray}
a_m=\frac{2g^2}{(b-v)^2+3g^2}
\end{eqnarray}
Haciendo un análisis similar al que se hizo para dos qubits (sección \ref{sec-dosqbits}) se ve que $a_m\ge1/2$ para $(b-v)^2/g^2<1$ y por ende existe un tiempo ($t_m=\lambda^{-1}\arcsin((2a_m)^{-1/2})$) para el cual se llega a un estado máximamente entrelazado. En cambio cuando $(b-v)^2/g^2>1$ nunca se a llega generar un estado máximamente entrelazado dado que $a(t)<1/2$ para todo $t$. El comportamiento del entrelazamiento de uno con el resto es por lo tanto igual que en el caso de dos qubits (fig \ref{fig:dosqbits}) pero la meseta esta desplazada en $v$ (limites dados por $(b-v)^2/g^2<1$) y cae asintóticamente  para $b\gg (v,g)$ como $C_1\approx2\sqrt2g/b$. \\
\begin{figure}[h]
\begin{center}
\epsfig{file=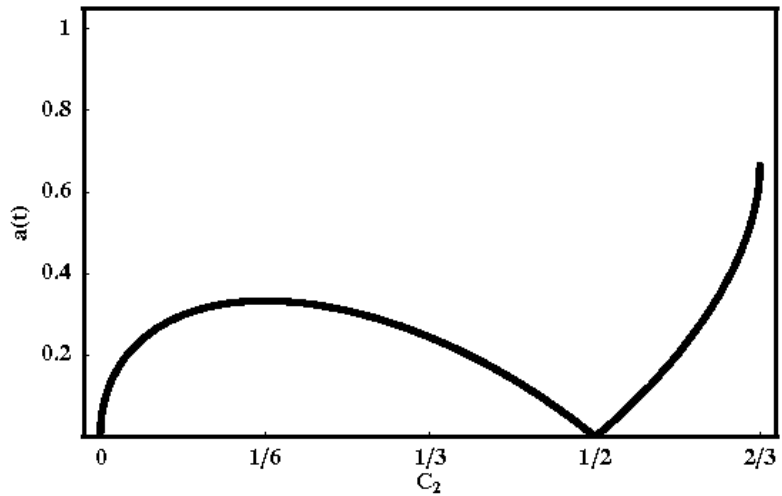, angle=0, width=300 pt}
\caption{\small  Entrelazamiento de pares $C_2$ como función de $a(t)$ para tres qubits.}\label{fig:tresqbits.concu}
\end{center}
\end{figure}

\begin{figure}[htbp]
  \begin{center}
    \mbox{
      \scalebox{0.8}{\epsfig{file=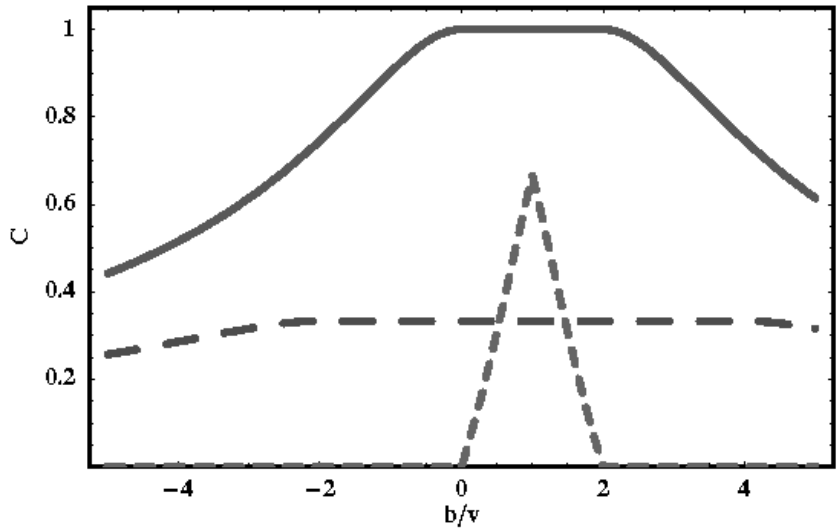}}
    }
    \caption{\small Entrelazamiento máximo de uno con el resto $C_1$ (línea sólida),  de pares $C_2$ con paridad positiva (línea a trazos largos) y de pares $C_2$ con paridad negativa (línea a trazos cortos) como función de  $b/v$ para $g=1$.}

    \label{fig:tresqbits}
  \end{center}
\end{figure}

El entrelazamiento de pares, al tratarse de más de dos qubits, difiere del de uno con el resto. Expresado como función de $a(t)$ el entrelazamiento de pares para este sistema es 
\begin{equation}
C_2=\left|a(t)-\sqrt{[2-3a(t)]a(t)}\right|
\end{equation}
La figura \ref{fig:tresqbits.concu} muestra esta función para $0<a(t)\le2/3$, que es el rango de valores que puede tomar $a(t)$. Obsérvese primero que nunca llegan a haber estados máximamente entrelazados $C_2<1\;\forall a(t)$. Hay dos regiones que se distinguen por la anulación de $C_2$ en $a(t)=1/2$. Para $a(t)$ por debajo de este límite los estados de Bell responsables del entrelazamiento (ver sección \ref{sec-concurrencia}) son los que tienen la misma paridad que el estado inicial y el máximo de esta parte se da cuando $a(t)=1/6$ donde $C_2=1/3$. Para valores de $a(t)>1/2$ hay tiempos en los cuales se generan estados entrelazados con paridad opuesta a la del estado inicial, aquí el máximo se da para $a(t)=2/3$ donde $C_2=2/3$ que es justamente el entrelazamiento de pares del estado de Werner. Es posible generar estos estados de paridad opuesta sólo cuando se esta cerca de la resonancia, es decir cuando $(b-v)^2/g^2<1$.\\
En la figura \ref{fig:tresqbits} se muestran los entrelazamientos mencionados. El entrelazamiento de uno con el resto, que tiene la misma forma que para el caso de dos qubits, el entrelazamiento de pares con paridad positiva que tiene un una meseta para $|(b-v)/g|<3$ y luego cae asintóticamente para $b\gg(v,g)$ como $C_2\approx\sqrt2g/|b|$ y el entrelazamiento de pares con paridad negativa que tiene un pico pronunciado centrado en $b=v$ de altura 2/3 y se anula para $|(b-v)/g|>1$.

\begin{figure}[h]
  \begin{center}
    \mbox{
      \subfigure[$(b-v)/g=0$]{\scalebox{0.5}{\epsfig{file=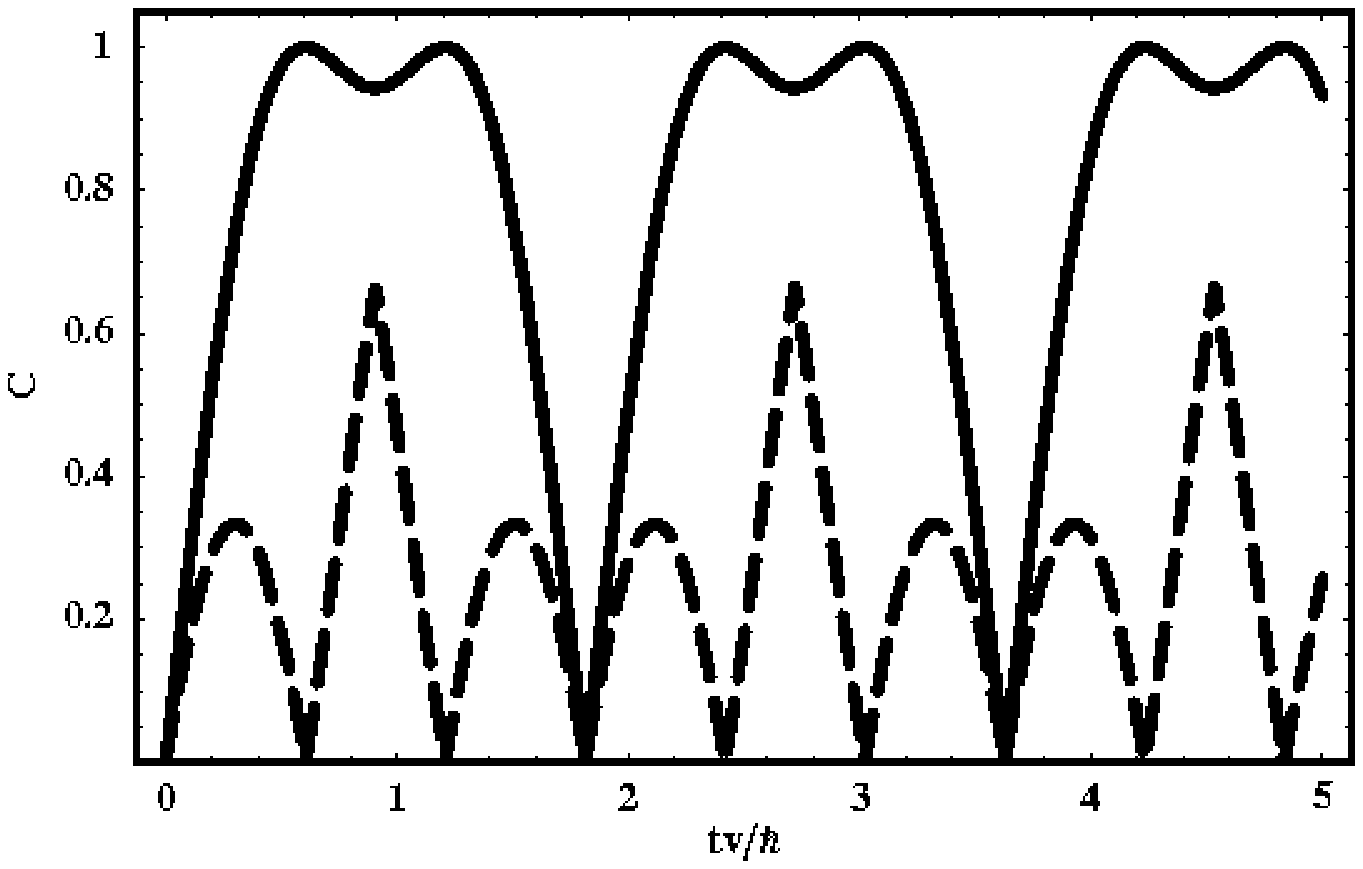}}\label{fig:ttresqbits0} }\quad
      \subfigure[$(b-v)/g=1$]{\scalebox{0.5}{\epsfig{file=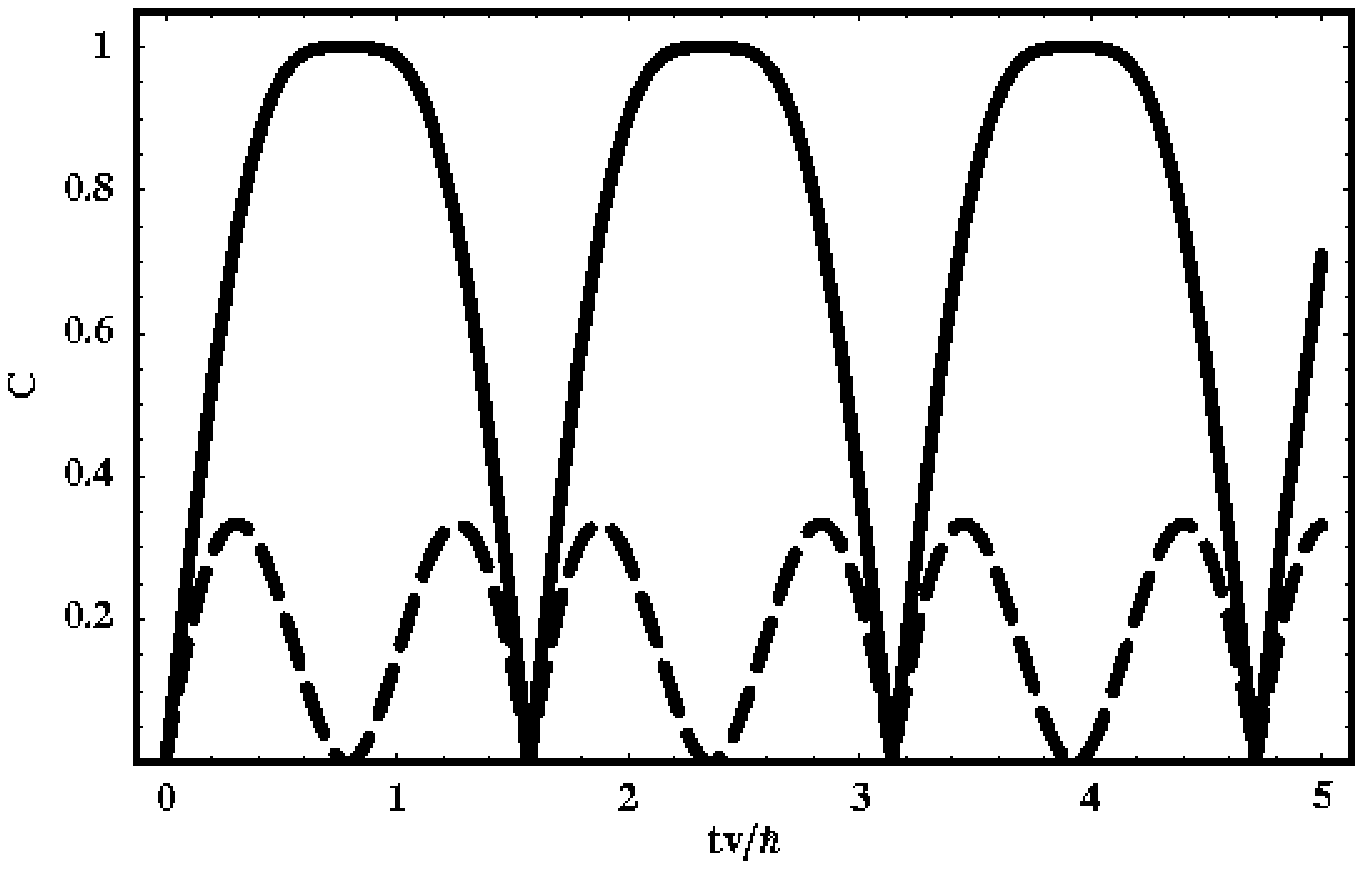}}\label{fig:ttresqbits1}}
      }
    \mbox{
      \subfigure[$(b-v)/g=2$]{\scalebox{0.5}{\epsfig{file=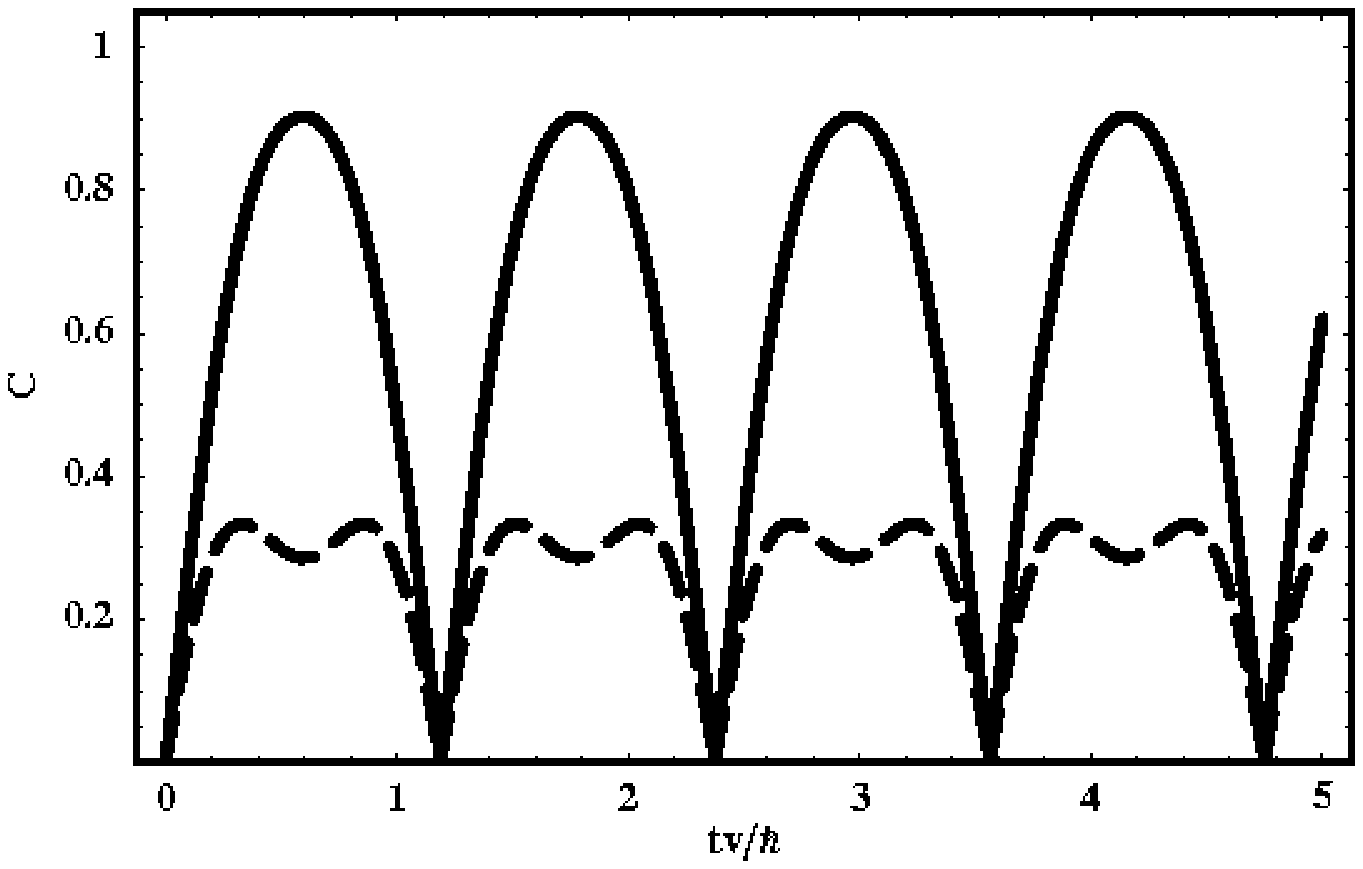}}\label{fig:ttresqbits2}} \quad
      \subfigure[$(b-v)/g=5$]{\scalebox{0.5}{\epsfig{file=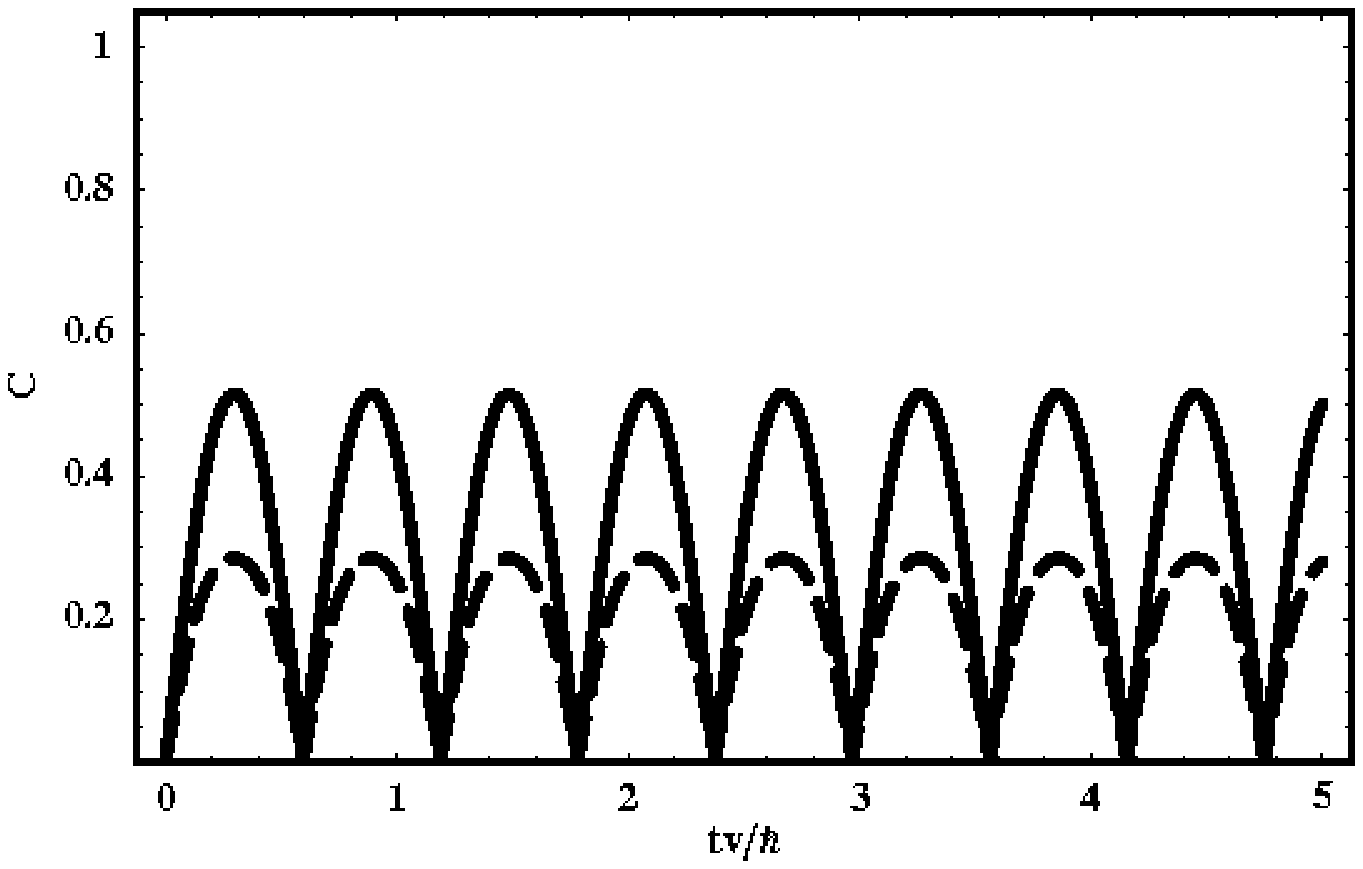}}\label{fig:ttresqbits5}} 
      }
    \caption{\small Entrelazamiento de uno con el resto (líneas sólidas) y entre pares (líneas punteadas) en función del tiempo en unidades de $\hbar/v$ para tres qubits y distintos valores de $(b-v)/g$.}
    \label{fig:ttresqbits}
  \end{center}
\end{figure}

A continuación mostramos la evolución temporal (fig \ref{fig:ttresqbits}) para cada una de las zonas. En la resonancia (fig \ref{fig:ttresqbits0}) se ven las características más relevantes del sistema: los picos de la concurrencia que corresponden a estados con paridad positiva (picos suaves) y negativa (picos agudos) y el entrelazamiento de uno con el resto que llega a la saturación en tiempos distintos a los que aparecen los picos de la concurrencia. En el punto crítico (fig \ref{fig:ttresqbits1}) , es decir donde $(b-v)/g=1$, el entrelazamiento ``se queda'' saturado por un tiempo (como en el caso de dos qubits) pero desaparece el pico de la concurrencia para estados con paridad negativa. A medida que los parámetros se alejan de la resonancia los picos de la concurrencia colapsan (fig \ref{fig:ttresqbits2}) y finalmente el comportamiento es similar al caso de dos qubits, el entrelazamiento de uno con el resto sigue a la concurrencia (fig \ref{fig:ttresqbits5}), pero siempre la concurrencia es menor.

\chapter[Cadenas XY largas]{Transformaciones Canónicas: Entrelazamiento en cadenas XY largas}

\subsection{Introducción}

En este capitulo resolvemos el problema de la evolución temporal de la concurrencia de pares y de uno con el resto para el sistema descripto por el Hamiltoniano del capítulo \ref{chap-sist} para una cantidad \emph{impar} arbitraria  de qubits.

El procedimiento\cite{LSM1961} consiste en hacer una transformada de Jordan-Wig\-ner\cite{JW1928} que mapea los operadores de espín $S_i^+$, $S_i^-$ a operadores puramente fermiónicos $c_i\da$,$c_i$ y luego diagonalizar el Hamiltoniano mediante una Transformación Canónica, que conserva las relaciones de conmutación.  Una transformación canónica puede ser descompuesta en tres partes\cite{RiSh1941}: una transformación unitaria que mapea los operadores $c_i$ en operadores $d_i$ del tipo $d_i=\sum_jM_{ij}c_i$, una trasformada especial de Bogoliubov (o transformada BCS) que mapea los operadores $d_i$ en operadores de cuasipartículas $a_i$ y una última transformación unitaria entre los operadores $a_i$. La primer transformación unitaria la descompongo en dos partes: un cambio de fase dependiente de la paridad elegida para eliminar el término no cuadrático que introduce la transformada de Jordan-Wigner con condiciones cíclicas y una transformada de Fourier discreta que lleva al Hamiltoniano a una forma canónica. Finalmente hago la transformada BCS que deja el Hamiltoniano en una forma diagonal de modo que la última transformación unitaria no es necesaria. Una vez diagonalizado el Hamiltoniano calculo exactamente los valores medios en función del tiempo necesarios para dar las formas explícitas de las concurrencias (ver ecuaciones \ref{ec.valores_medios_concu}).\\
Para el caso de tres qubits los resultados obtenidos coinciden exactamente con los de la sección \ref{sec.tres_qbits}. \\
Finalmente se hace un análisis de la dependencia funcional del máximo de las concurrencias con el campo magnético $b$ para distintos rangos de $v,g$ (parámetro de ``hopping'', anisotropía de la interacción).
\section{Diagonalización}
\subsection{Transformación de Jordan-Wigner}
Partimos del mismo Hamiltoniano de espines $1/2$ en un campo magnético que usamos en la sección anterior.
\begin{equation}
H=\sum_{i=1}^n \bigl[  b S_i^z - (vS_i^+S_{i+1}^- +gS_i^+S_{i+1}^+ +h.c.)\bigr] 
\end{equation}
Las relaciones de conmutación de los operadores de subida y bajada ($S_i^+$ , $S_i^-$) son
en parte fermiónicas
\begin{equation}\{S_i^-,S_i^+\}=1\;;\;(S_i^+)^2=(S_i^-)^2=0\end{equation}
y en parte bosónicas
\begin{equation}[S_i^+,S_j^-]=[S_i^+,S_j^+]=[S_i^-,S_j^-]=0\;;\;i\neq j\end{equation}
Para poder diagonalizar el Hamiltoniano con transformaciones canónicas (Fourier y Bogoliubov) como haremos luego  es preciso que las relaciones de conmutación de los operadores sean fermiónicos o bosónicas, pero no una mezcla como es el caso de $S_i^+$ , $S_i^-$. Para llevar estos operadores a una forma puramente fermiónica hacemos ahora una transformación de Jordan-Wigner \cite{JW1928} que mapea los operadores $S^+$ y $S^-$ en operadores $c^\dagger_i$,$c_i$ puramente fermiónicos.
\begin{eqnarray}
c_i=exp\bigl[\pi i \sum_{j=1}^{i-1}S^+_jS^-_j]S^-_i&& 
S_i^-=exp\bigl[- \pi i \sum_{j=1}^{i-1}c^\dagger_jc_j]c_i\\
c_i^\dagger=S^+_iexp\bigl[-\pi i \sum_{j=1}^{i-1}S^+_jS^-_j]&&
S_i^+=c^\dagger_i exp\bigl[\pi i \sum_{j=1}^{i-1}c^\dagger_jc_j]
\end{eqnarray}
donde se tiene que:
\begin{equation}\{
c_i,c_j^\dagger\}=\delta_{ij},\;\{c_i,c_j\}=\{c_i^\dagger,c_j^\dagger\}=0
\end{equation}

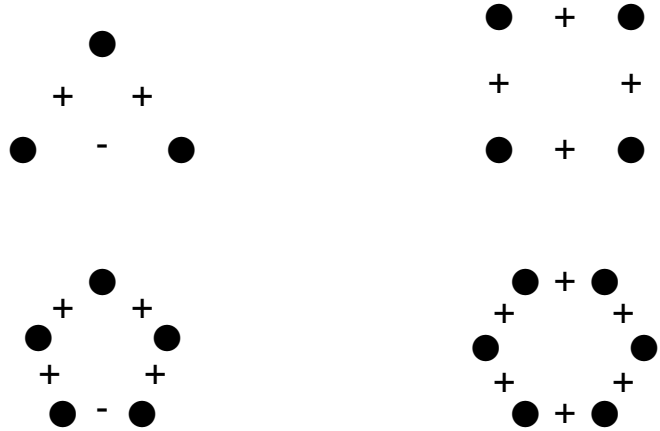
\begin{figure}
\begin{picture}(200,200)(-80,-100)
\put(30,20){\makebox(0,0){-}}
\put(15,40){\makebox(0,0){+}}
\put(45,40){\makebox(0,0){+}}
\put(0,20){\circle*{10}}
\put(60,20){\circle*{10}}
\put(30,60){\circle*{10}}

\put(180,20){\circle*{10}}
\put(230,20){\circle*{10}}
\put(180,70){\circle*{10}}
\put(230,70){\circle*{10}}
\put(205,20){\makebox(0,0){+}}
\put(180,45){\makebox(0,0){+}}
\put(205,70){\makebox(0,0){+}}
\put(230,45){\makebox(0,0){+}}

\put(15,-80){\circle*{10}}
\put(5.7,-51.4){\circle*{10}}
\put(54.5,-51.4){\circle*{10}}
\put(45,-80){\circle*{10}}
\put(30,-30){\circle*{10}}
\put(30,-80){\makebox(0,0){-}}
\put(50,-65){\makebox(0,0){+}}
\put(10,-65){\makebox(0,0){+}}
\put(15,-40){\makebox(0,0){+}}
\put(45,-40){\makebox(0,0){+}}

\put(190,-80){\circle*{10}}
\put(220,-80){\circle*{10}}
\put(175,-55){\circle*{10}}
\put(235,-55){\circle*{10}}
\put(190,-30){\circle*{10}}
\put(220,-30){\circle*{10}}
\put(182,-68){\makebox(0,0){+}}
\put(227,-68){\makebox(0,0){+}}
\put(182,-42){\makebox(0,0){+}}
\put(227,-42){\makebox(0,0){+}}
\put(205,-80){\makebox(0,0){+}}
\put(205,-30){\makebox(0,0){+}}
\end{picture}
\caption{\small Signo de la integración entre espines para sistemas de 3, 4, 5 y 6 espines. En los casos impares queda una interacción con signo distinto.}\label{fig:cambio.signo.interac}
\end{figure}

Para poder escribir el Hamiltoniano en términos de estos nuevos operadores vemos que de las definiciones de las transformaciones se desprende que:
\begin{equation}c_i\da c_i=S_i^+ S_i^-\end{equation}
Además, dado que $c_i\da c_i$ es un número de ocupación que puede ser 0 o 1 se tiene que:
\begin{equation}exp[\pi i c_i\da c_i]=exp[-\pi i c_i\da c_i]\end{equation}
Además para $i=1,2,\ldots,N-1$ se tiene que:
\begin{eqnarray}c_i\da c_{i+1}=S_i^+ S_{i+1}^-&;&
c_i\da c_{i+1}\da=S_i^+ S_{i+1}^+\end{eqnarray}
de modo que, para el caso de extremos libres, el Hamiltoniano es
\begin{equation}
H=\sum_{i=1}^{n-1} \bigl[  b (c_i\da c_i-\frac12) - (vc_i\da c_{i+1} +gc_i\da c_{i+1}\da +h.c.)\bigr] 
\end{equation}
El caso que nos interesa resolver, sin embargo, es el de la cadena cíclica. Debemos entonces tener en cuenta los términos
\begin{subequations}\begin{eqnarray}
S_n^+ S_{1}^-&=-c_n\da c_{1}P&\neq c_n\da c_{1}\\
S_n^+ S_{1}^+&=-c_n\da c_{1}\da P&\neq c_n\da c_{1}\da
\end{eqnarray}\end{subequations}
donde $P$ es el operador paridad de espín (apéndice \ref{chap-par}). De este modo se obtiene el 
Hamiltoniano para el caso cíclico.
\begin{eqnarray}
H&=&\sum_{i=1}^n \bigl[  b (c_i\da c_i-\frac12) - (vc_i\da c_{i+1} +gc_i\da c_{i+1}\da +h.c.)\bigr]\nonumber\\
&&+\bigl[ vc_n\da c_{1} +gc_n\da c_{1}\da +h.c.)\bigr](P+1)
\end{eqnarray}
El desarrollo hasta aquí es general. De ahora en adelante empezaré a hacer suposiciones sobre el sistema que voy a estudiar.\\
Como se mencionó en la sección \ref{sec-parham} tanto el estado inicial como el Hamiltoniano conmutan con $P$ y el estado inicial tiene paridad positiva entonces se puede reemplazar $P$ por su valor ya que este permanecerá invariante. Se tiene entonces:
\begin{eqnarray}
H&=&\sum_{i=1}^n\bigl[  b (c_i\da c_i-\frac12) - (vc_i\da c_{i+1} +gc_i\da c_{i+1}\da +h.c.)\bigr]\nonumber\\
&&+2\bigl[ vc_n\da c_{1} +gc_n\da c_{1}\da +h.c.)\bigr]
\end{eqnarray}
El término que no está en la sumatoria tiene signo opuesto a los términos de interacción en la sumatoria. En pos de llevar este Hamiltoniano a una forma más simple, que se pueda escribir como una simple sumatoria, hago la primer transformación lineal $c_i\rightarrow(-1)^ic_i$ que cambia los signos de las interacciones. En la figura \ref{fig:cambio.signo.interac} se muestra esquemáticamente como quedan las interacciones de los términos de la sumatoria. Para $n$ par cambian todos los términos de signo pero para $n$ impar el término que tiene las interacciones entre el el primer y el último espín no cambia. En este último caso -$n$ impar-  puede escribirse entonces el Hamiltoniano\footnote{Este Hamiltoniano también describe adecuadamente cadenas con $n$ par pero estado inicial impar.}
\begin{eqnarray}
H&=&\sum_{i=1}^n\bigl[  b (c_i\da c_i-\frac12) + (vc_i\da c_{i+1} +gc_i\da c_{i+1}\da +h.c.)\bigr]
\end{eqnarray}

\subsection{Transformación de Fourier Discreta}
Para diagonalizar este Hamiltoniano se realiza primero una transformada de Fourier discreta de los operadores $c\da_i, c_i$ dada por
\begin{eqnarray}
c_j\da=\frac{e^{-i \pi/4}}{\sqrt{n}}\sum_{k=1}^n d_k\da e^{i w_k j} &;&
w_k=2 \pi k /n
\end{eqnarray}
Se obtiene entonces, para el caso de una cantidad impar de qubits:
\begin{eqnarray}
H=\sum_{i=1}^n \left[  (b+2v\cos{w_i}) d_i\da d_i+ g\sin{w_i}(d_i\da d_{n-i}\da+h.c.)\right]-\frac{n b}2
\end{eqnarray}
\subsection{Transformación de Bogoliubov}
Para terminar la diagonalización se propone una transformada especial de Bogoliubov de la forma:
\begin{subequations}
\begin{eqnarray}
d_j\da=u_ja_j\da+v_ja_{n-j}\\
d_{n-j}\da=u_ja_{n-j}\da-v_ja_{j}
\end{eqnarray}
\end{subequations}
y se lleva a H a la forma diagonal (ver apéndice \ref{ap-bogo.espe})
\begin{subequations}\begin{eqnarray}
\label{ec.hamdiag}
H&=&\sum_j \bigl[\lambda_j  (a_j\da a_j -\frac12)\bigr]+
cte\\
 \lambda_j&=& \sqrt{(b+2v\cos{w_j})^2+4g^2 \sin^2{w_j}}
\end{eqnarray}\end{subequations}
esto se logra eligiendo $(u_j^2,v_j^2)=\frac12[1\pm(b+2v\cos{w_j})/\lambda_j]$. La constante aditiva en el Hamiltoniano no juega ningún papel en el resultado final.
\subsection{Evolución Temporal y Valores Medios}
Dado que el Hamiltoniano es diagonal en la base de operadores de cuasipartículas de Bogoliubov es fácil calcular la evolución temporal de los valores medios en la representación de Heisenberg. Esto es
\begin{eqnarray}
\braketo{O(t)}=Tr(\rho_0 O(t)) &;&\frac{d O(t)}{dt}=i[H,O(t)]
\end{eqnarray}
De modo que usando \ref{ec.hamdiag} y teniendo en cuenta que los $a_i$,$a_i\da$ son fermiónicos,  para $a_j$ se tiene
\begin{eqnarray}
\frac{da_j}{dt}
&=&i[H,a_j]\nonumber\\
&=&-i\lambda_ja_j
\end{eqnarray}
y análogamente para $a_j\da$. Integrando elementalmente se tiene
\begin{eqnarray}
a_j\da(t)=e^{i\lambda_j t}a_j\da(0) &\mbox{y}& a_j(t)=e^{-i\lambda_j t}a_j(0) 
\label{ec:deptempcuasi}
\end{eqnarray}
Para poder construir las medidas de entrelazamiento es necesario conocer los valores medios de $S^z_{i}$, $S^+_{i}S^-_{i+1}$ , $S^+_{i}S^+_{i+1}$ y $S^z_{i}S^z_{i+1}$.\\
Los primeros tres operadores pueden ser escritos como función de los operadores de cuasipartículas de Bogoliubov (dejando su dependencia temporal implícita) aplicando las transformadas y considerando que como la cadena es cíclica $O_j=\sum_{i=1}^{n}O_i/n$.
\begin{eqnarray}
S^z_{i}+\frac12&=&\frac1n\sum_{j=1}^n\left[
u_j^2 a_j\da a_j + v_j^2a_{n-j}a_{n-j}\da+u_jv_j(a_j\da a_{n-j}\da+a_{n-j}a_j)\right]\nonumber\\
S^+_{i}S^-_{i+1}&=&\frac1n\sum_{j=1}^n e^{-iw_j}\left[
u_j^2 a_j\da a_j + v_j^2a_{n-j}a_{n-j}\da+u_jv_j(a_j\da a_{n-j}\da+a_{n-j}a_n)\right]\nonumber\\
S^+_{i}S^+_{i+1}&=&\frac1n\sum_{j=1}^n e^{-iw_{n-j}}\left[
u_j^2 a_j\da a_{n-j}\da -v_j^2 a_{n-j} a_j +u_jv_j(a_{n-j} a_{n-j}\da-a_j\da a_j)\right]\nonumber\\\label{ec:operadores.bogo}
\end{eqnarray}
El valor medio operador restante puede ser escrito en función de los primeros (Teorema de Wick\cite{RiSh1941}) como:
\begin{eqnarray}
\braketo{S^z_{i}S^z_{i+1}}=\braketo{S^z_{i}}^2
+\abs{\braketo{S^+_{i}S^+_{i+1}}}^2
-\abs{\braketo{S^+_{i}S^-_{i+1}}}^2
\end{eqnarray}
Para construir los valores medios $\braketo{S^z_{i}}$, $\braketo{S^+_{i}S^-_{i+1}}$ y $\braketo{S^+_{i}S^+_{i+1}}$, como  se ve en las ecuaciones \ref{ec:operadores.bogo}, es necesario primero conocer $\braketo{a\da_ja_j}$, $\braketo{a_{n-j}a\da_{n-j}}$ y $\braketo{a_j\da a_{n-j}\da}$. Cómo la dependencia temporal de los operadores de cuasipartículas es simplemente una fase (ec. \ref{ec:deptempcuasi}) se puede hacer el cálculo de estos valores medios a $t=0$ y luego multiplicar por la fase al resumar para armar las expresiones \ref{ec:operadores.bogo}. 
Además, es crucial notar que estos son valores medios de operadores de cuasipartículas medidos con respecto al estado inicial, que no es más que el vacío de los operadores fermiónicos de Jordan-Wigner ($c_i\da,c_i$). Para poder calcularlos es
necesario expresar los operadores de Bogoliubov como función de los de Jordan-Wigner mediante las transformaciones inversas de Fourier y de Bogoliubov. Estas son
\begin{eqnarray}
a_j\da=u_jd_j\da-v_jd_{n-j}&;& 	
a_{n-j}\da=ud_{n-j}\da+vd_{j}
\end{eqnarray}
y
\begin{eqnarray}
d_j\da&=&\frac{e^{i \pi/4}}{\sqrt{n}}\sum_{k=1}^n c_k\da e^{-i w_k j}
\end{eqnarray}
De modo que que se obtiene:
\begin{subequations}
\begin{eqnarray}
\braketo{a\da_ja_j}&=&v_j^2\\
\braketo{a_{n-j}a_{n-j}\da}&=&u_j^2\\
\braketo{a_j\da a_{n-j}\da}&=& -u_jv_je^{2i\lambda_jt} =\braketo{a_{n-j}a_{j}}^*
\end{eqnarray}
\end{subequations}
Juntando todo se obtienen los valores medios deseados.
\begin{flushleft}
\begin{subequations}
\label{eq:bogo.res}
\begin{eqnarray}
\braketo{S^z_{i}}&=& 
\frac1n \sum_{j=1}^{n-1}\frac{4 g^2 \sin^2{w_j}}{\lambda_j^2}\sin^2{\lambda_j t}-1/2  
\label{eq:bogo.sz}\\
\braketo{S^+_{i}S^-_{i+1}}&=&
\frac1n \sum_{j=1}^{n-1}\frac{4 g^2 \sin^2{w_j}\cos{w_j}}{\lambda_j^2}\sin^2{\lambda_j t}            \\
\braketo{S^+_{i}S^+_{i+1}}&=& 
-\frac{1}n \sum_{j=1}^{n-1}\frac{g \sin^2{w_j}}{\lambda_j}\bigl[2\frac{b+2v \cos{w_j}}{\lambda_j}\sin^2{\lambda_j t}+\nonumber\\
&&\hspace{7.3cm} i\sin{2\lambda_j t}\bigr]\\
\braketo{S^z_{i}S^z_{i+1}}&=&\braketo{S^z_{i}}^2
+\abs{\braketo{S^+_{i}S^+_{i+1}}}^2
-\abs{\braketo{S^+_{i}S^-_{i+1}}}^2
\end{eqnarray}
\end{subequations}
\end{flushleft}

\section{Entrelazamiento}
\subsection{Generalidades}
En este caso la función $a(t)$  que determina el entrelazamiento de uno con el resto es simplemente $\braketo{S^z_{i}}+1/2$ (ver ecuación \ref{eq:bogo.sz}). Analizaremos primero la forma de esta y daremos resultados generales y luego indicaremos generalidades sobre el entrelazamiento de pares que es una función mucho más complicada de la ecuaciones \ref{eq:bogo.res}.\\
La envolvente superior de $a(t)$ se obtiene haciendo\footnote{Sólo es posible que se de esta situacíón si las razones entre las frecuencias no son números racionales.} $\sin^2{\lambda_j t}=1\;\forall j$ 
y es
\begin{equation}
a_m= \frac1n \sum_{j=1}^{n-1}\frac{4 g^2 \sin^2{w_j}}
{(b+2v\cos{w_j})^2+4g^2 \sin^2{w_j}}
\end{equation}
Este resultado es una generalización natural de los anteriores. Es una suma de gausianas con centros en $b=-2v\cos{w_j}$ y de ancho $2g|\sin{w_j}|$. Para $n$ par la cantidad de picos distintos es $n/2+1$, mientras que para $n$ impar, lo que nos interesa ahora, hay $(n-1)/2$  picos. Como siempre, nos interesa saber cuando esta función es menor o mayor a $1/2$  ya que esto determina si se llega al entrelazamiento máximo o no.

\begin{figure}[h]
  \begin{center}
    \mbox{
      \subfigure[$g=1.0$] {\scalebox{0.5}{\epsfig{file=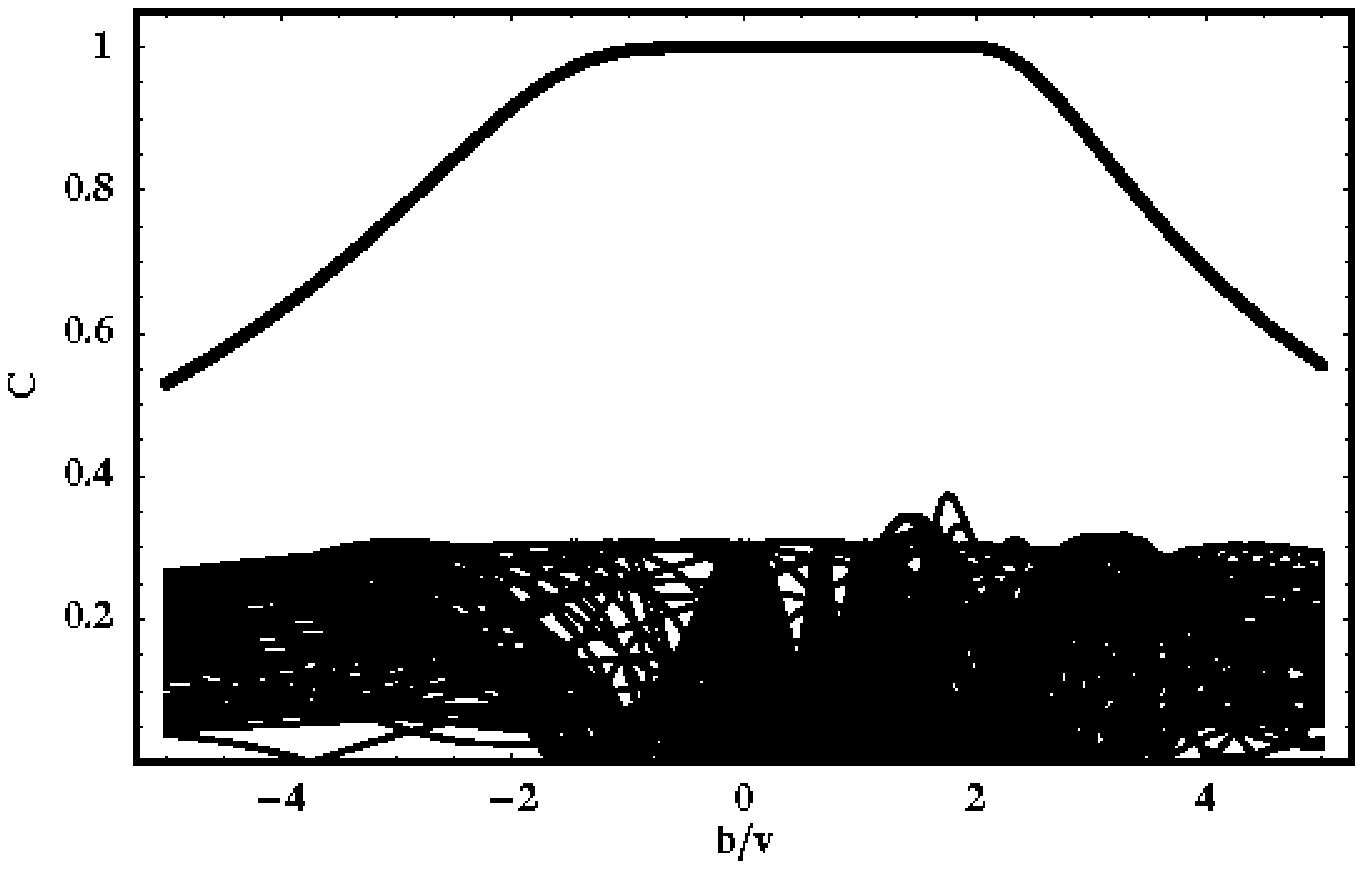}}\label{fig:bogon5.10} } \quad
      \subfigure[$g=0.5$] {\scalebox{0.5}{\epsfig{file=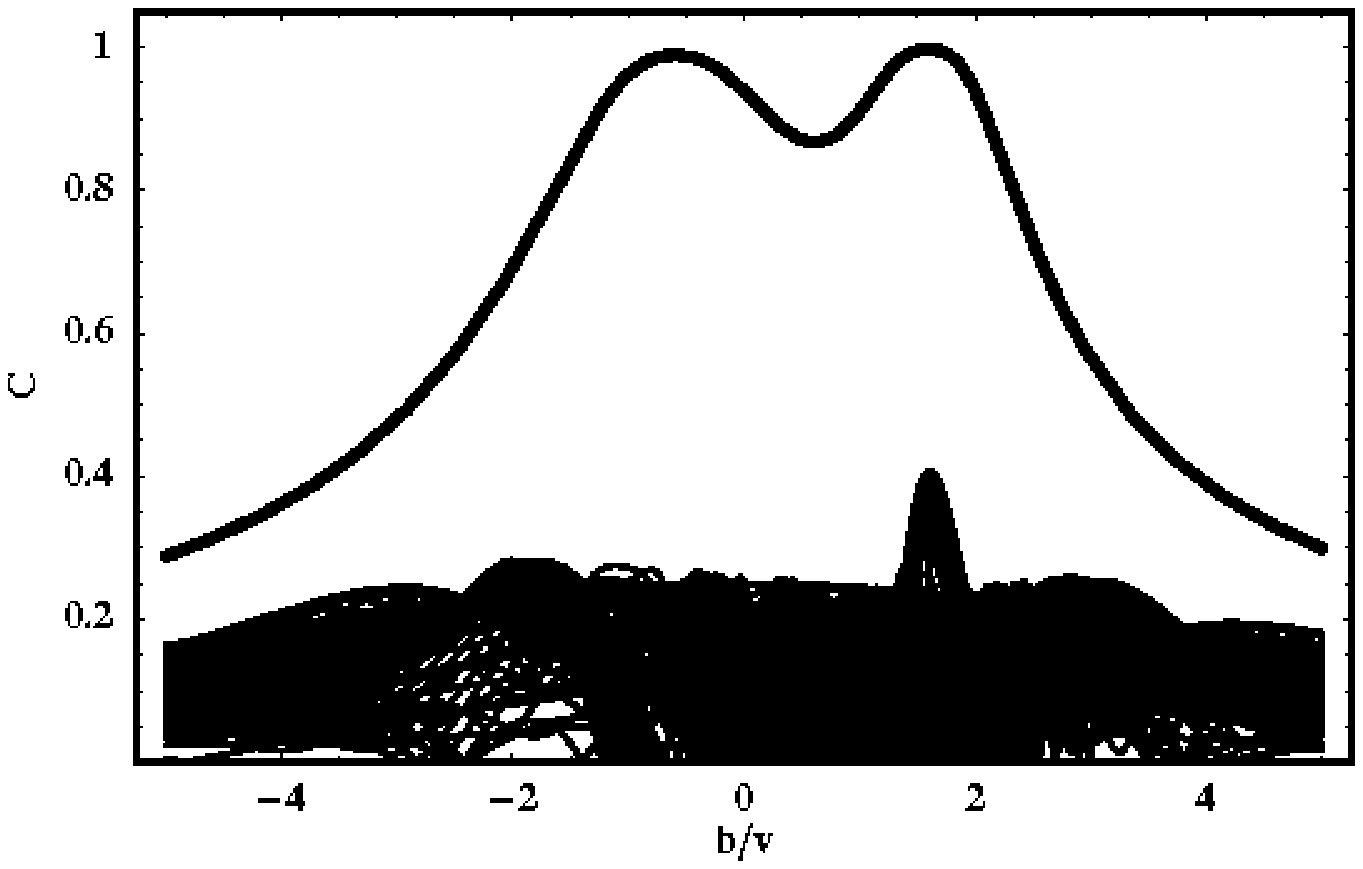}}\label{fig:bogon5.05}
	}
    } 
    \mbox{
      \subfigure[$g=0.3$] {\scalebox{0.5}{\epsfig{file=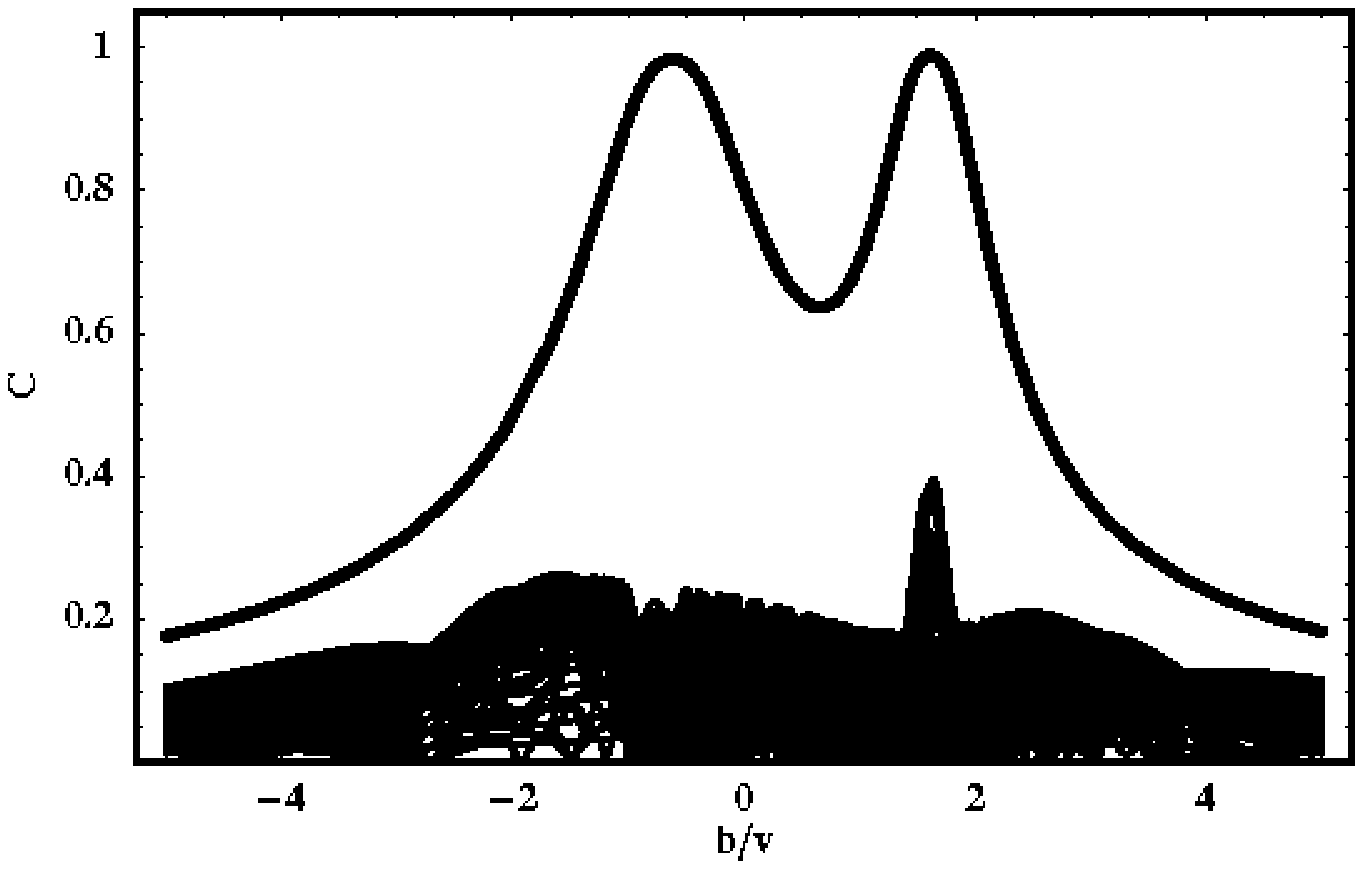}}\label{fig:bogon5.03}}\quad
      \subfigure[$g=0.1$] {\scalebox{0.5}{\epsfig{file=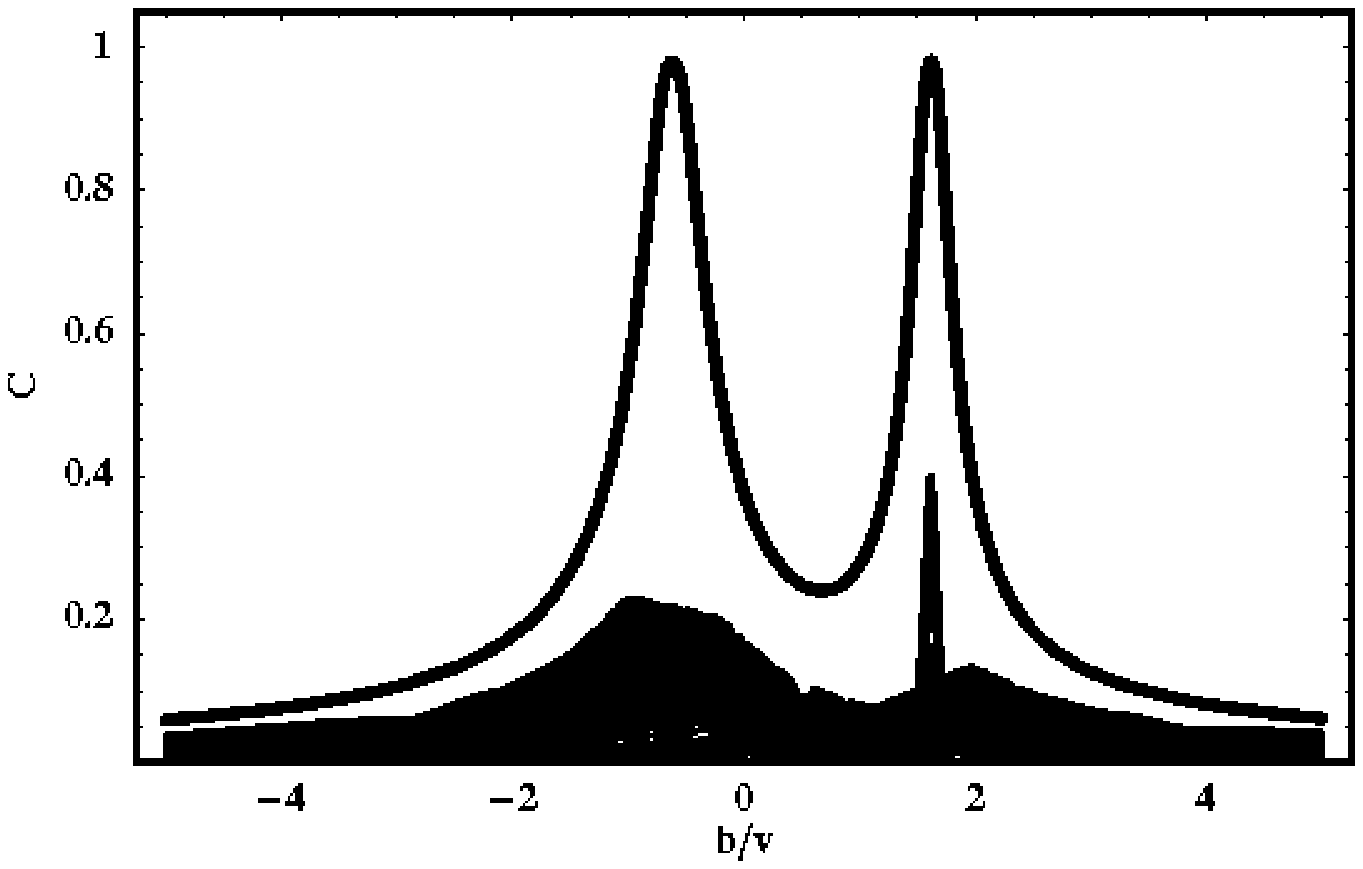}}\label{fig:bogon5.01}}
    }
    \caption{\small Entrelazamiento máximo de uno con el resto $C_1$ y de a pares $C_2$ para $n=5$ en función de $b/v$ para distintos valores de $g$ (en unidades de $v$). La línea gruesa es el entrelazamiento de uno con el resto $C_1$. Lo que se ve por debajo de esta es el entrelazamiento de pares $C_2$ para muchos tiempos distintos. }
    \label{fig:bogon5}
  \end{center}
\end{figure}

\begin{figure}[h]
  \begin{center}
    \mbox{
      \subfigure[ $g=1.0$] {\scalebox{0.5}{\epsfig{file=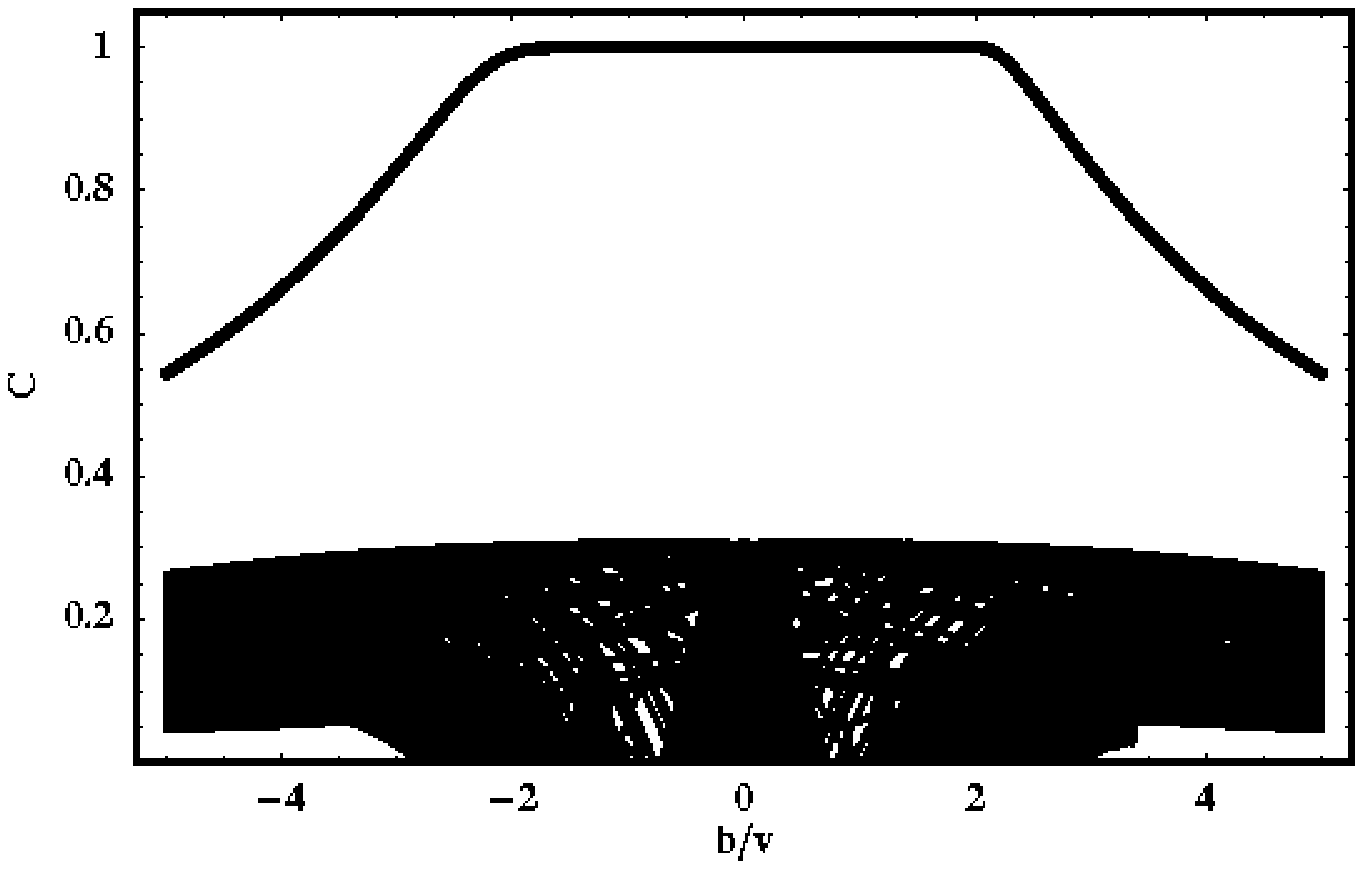}}\label{fig:bogon15.10} }\quad
      \subfigure[ $g=0.5$] {\scalebox{0.5}{\epsfig{file=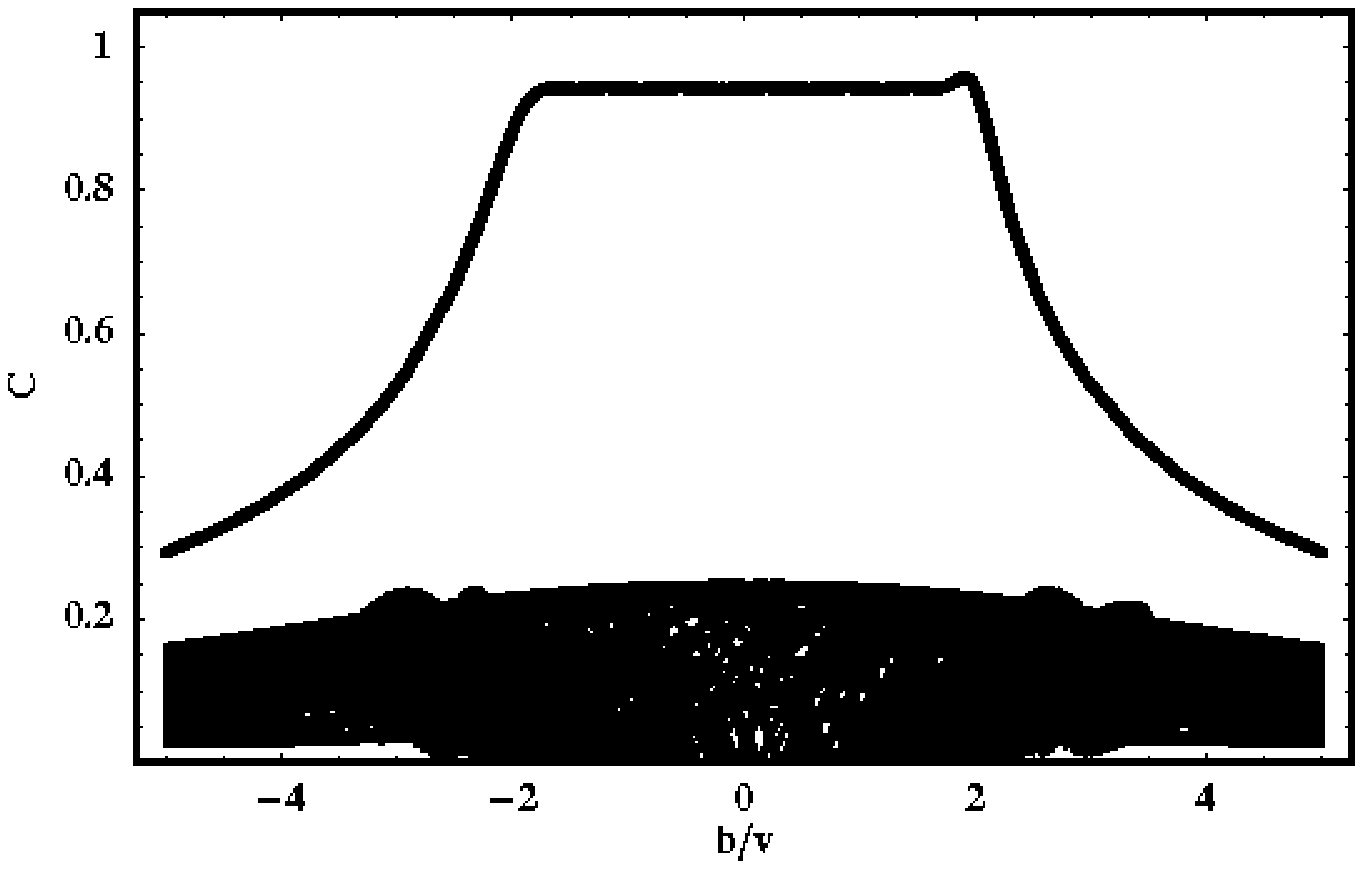}}\label{fig:bogon15.05}}
    }
    \mbox{
      \subfigure[ $g=0.3$] {\scalebox{0.5}{\epsfig{file=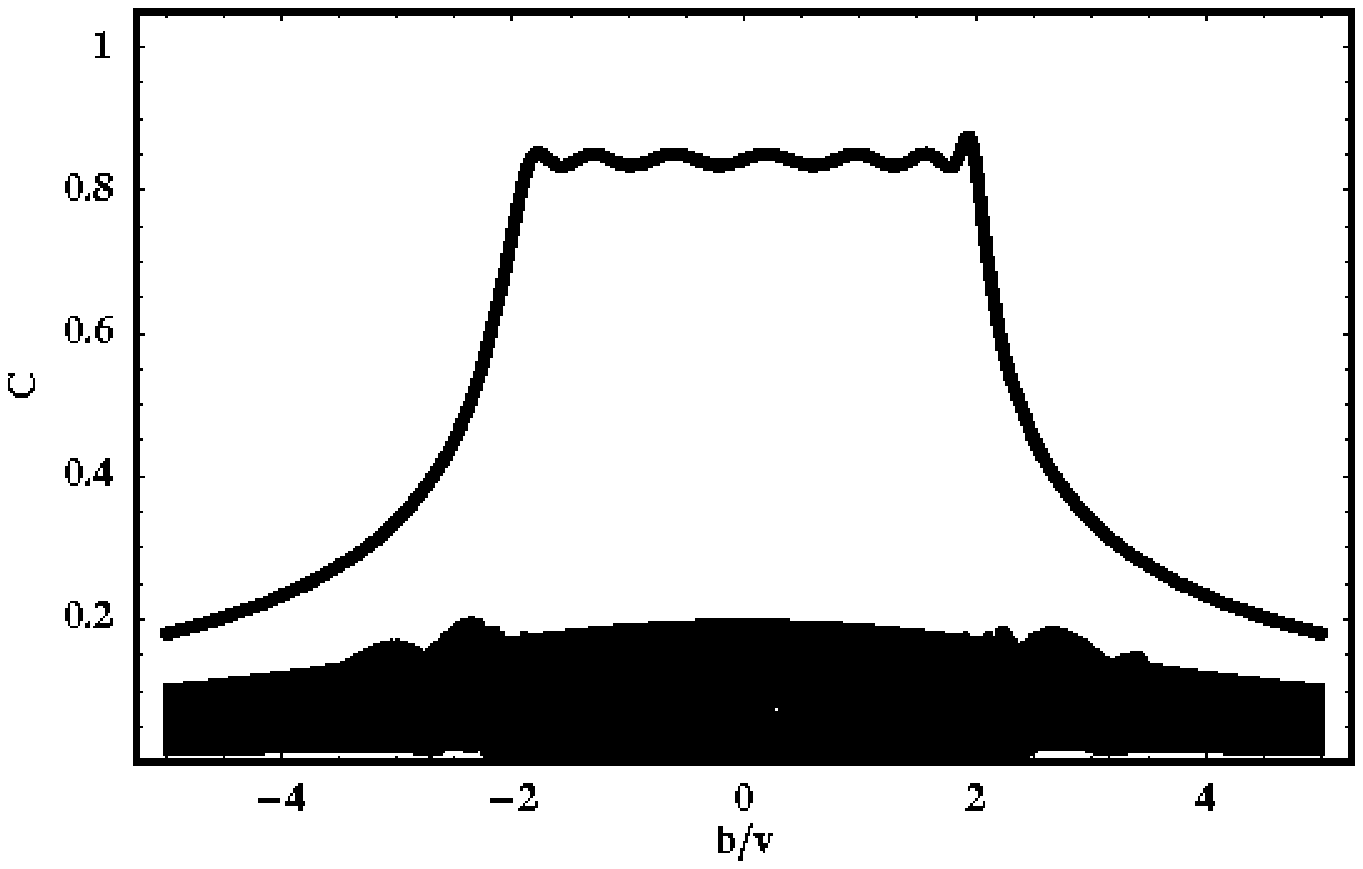}}\label{fig:bogon15.03}}\quad
      \subfigure[$g=0.1$] {\scalebox{0.5}{\epsfig{file=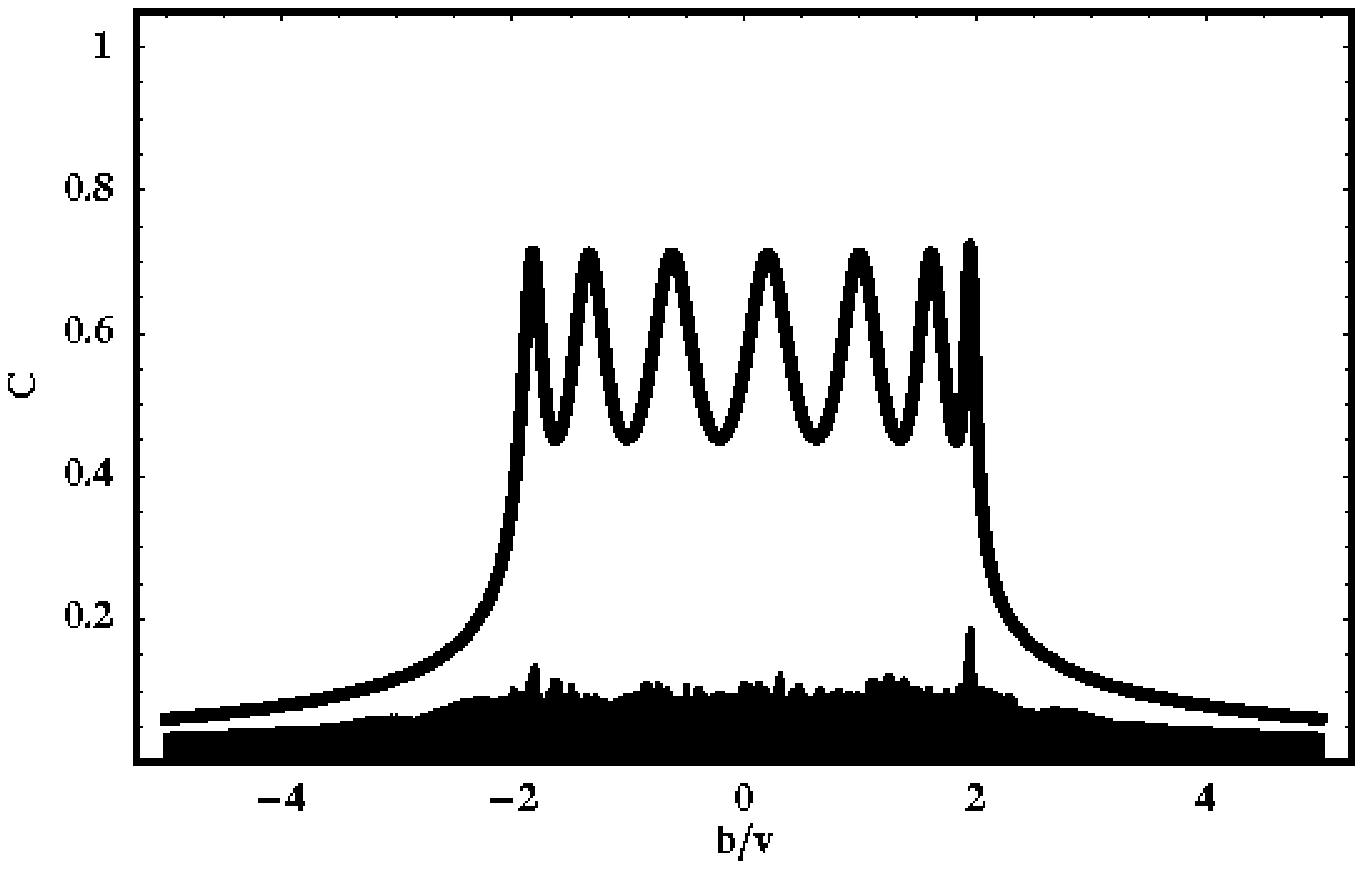}}\label{fig:bogon15.01}}
      }
    \caption{\small Entrelazamiento máximo de uno con el resto $C_1$ y de a pares $C_2$ para $n=15$ en función de $b/v$ para distintos valores de $g$. La línea gruesa es el entrelazamiento de uno con el resto $C_1$. Lo que se ve por debajo de esta es el entrelazamiento de pares $C_2$ para muchos tiempos distintos. }
    \label{fig:bogon15}
  \end{center}
\end{figure}

La condición para que un pico $j$ aislado llegue a un valor mayor a 1/2 es $(b+2v\cos{w_j})^2<4g^2 \sin^2{w_j}$ de lo que se puede deducir que si $v\lesssim g$ y $b\lesssim 2v$  se tiene $a(t)>1/2$. Estos resultados pueden constatarse comparando las figuras \ref{fig:bogon5.10} y \ref{fig:bogon15.10} con las \ref{fig:bogon5.05} y \ref{fig:bogon15.05} donde se  muestra el entrelazamiento de uno con el resto para  $g=v=1$ (hay saturación) y  $g=v/2=1$ (no hay saturación) correspondientemente, para dos sistemas de distinta cantidad de qubits.

Si $g\lesssim v$ no sólo se deja de llegar a la saturación sino que a medida que $g$ se hace más chico que $v$ se empiezan a distinguir los picos. La distancia entre dos picos consecutivos es $\delta b=2v|\cos{w_j}-\cos{w_{j+1}}|\approx2v|\frac{d\cos{w_j}}{d w_j}|\delta w_j=\frac{4\pi v}{n}|\sin{w_j}|$ y el ancho de un pico es $2g|\sin{w_j}|$. Podemos estimar que se verán picos claros cuando la separación entre ellos sea mayor que sus anchos; eso  es $g\lesssim2\pi v/n$. Esto puede constatarse con los gráficos \ref{fig:bogon5} y \ref{fig:bogon15}. Para el caso de $n=5$ se tiene $2\pi v/n\approx 1.26$ por lo que en todos los casos mostrados se deberían distinguir picos sin embargo no es así porque para $v=1$ la saturación borra los picos. Para $n=15$ se tiene $2\pi v/n\approx 0.42$ y como se ve sólo se resuelven los picos en los gráficos \ref{fig:bogon15.03} y \ref{fig:bogon15.01} que corresponden a $v=0.3$ y $v=0.1$ respectivamente; en \ref{fig:bogon15.10} y \ref{fig:bogon15.05} que corresponden a $v=1$ y $v=0.5$ no se resuelven picos.

En el límite asintótico de $b\gg (v,g)$ se obtiene $a_m\approx2g^2/b^2$ que es el mismo resultado que para tres qubits y por ende para cualquier $n$ impar, en este límite, $C_1\approx2\sqrt{2}g/|b|$.

En general el entrelazamiento de pares está muy por debajo del entrelazamiento de uno con el resto y lejos de la resonancia es prácticamente una función monótonamente decreciente del campo $b$; en el limite asintótico de $b\gg(v,g)$ se tiene $C_2\approx 2 g/b$. Sin embargo, para los casos en que hay resonancia la concurrencia también resuena y veremos más adelante que se pueden encontrar resultados interesantes.

\subsection{Entrelazamiento y Decoherencia}
\begin{figure}[t]
  \begin{center}
    \mbox{
      \subfigure[n=5]{\scalebox{0.7}[0.7]{\epsfig{file=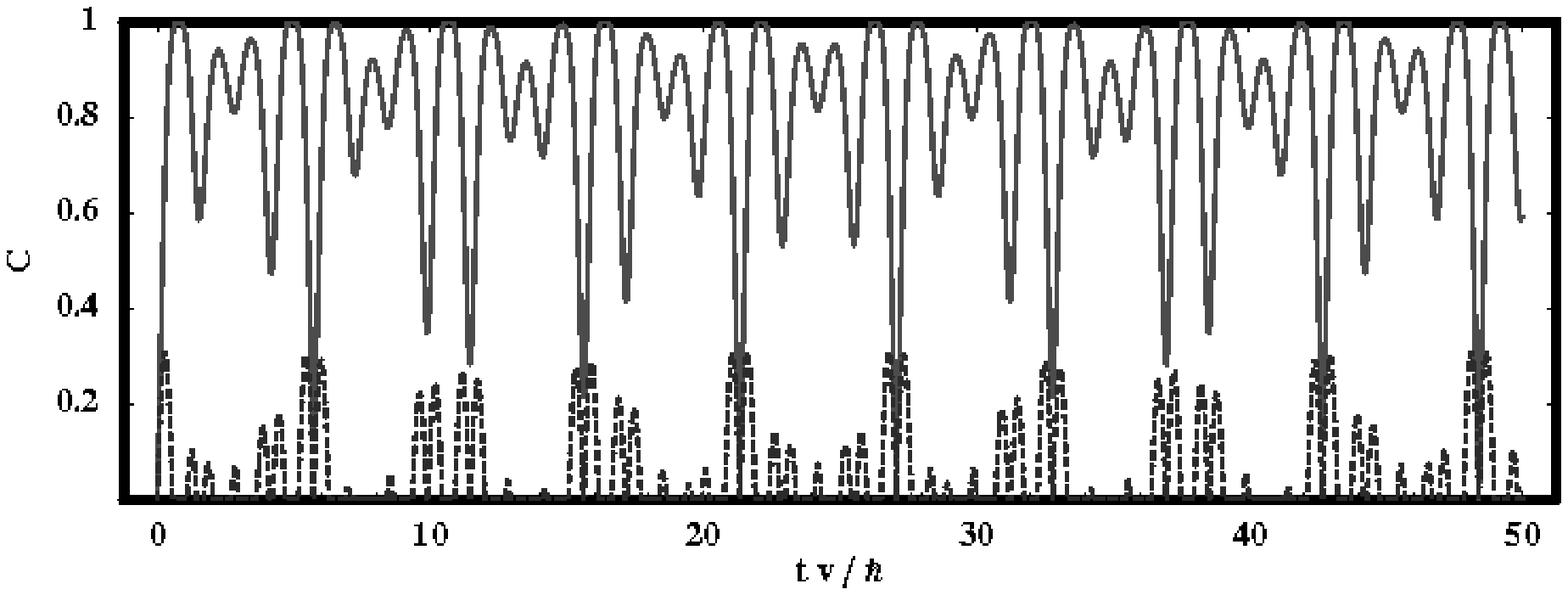}}}
    }
    \mbox{
      \subfigure[n=15]{\scalebox{0.7}[0.7]{\epsfig{file=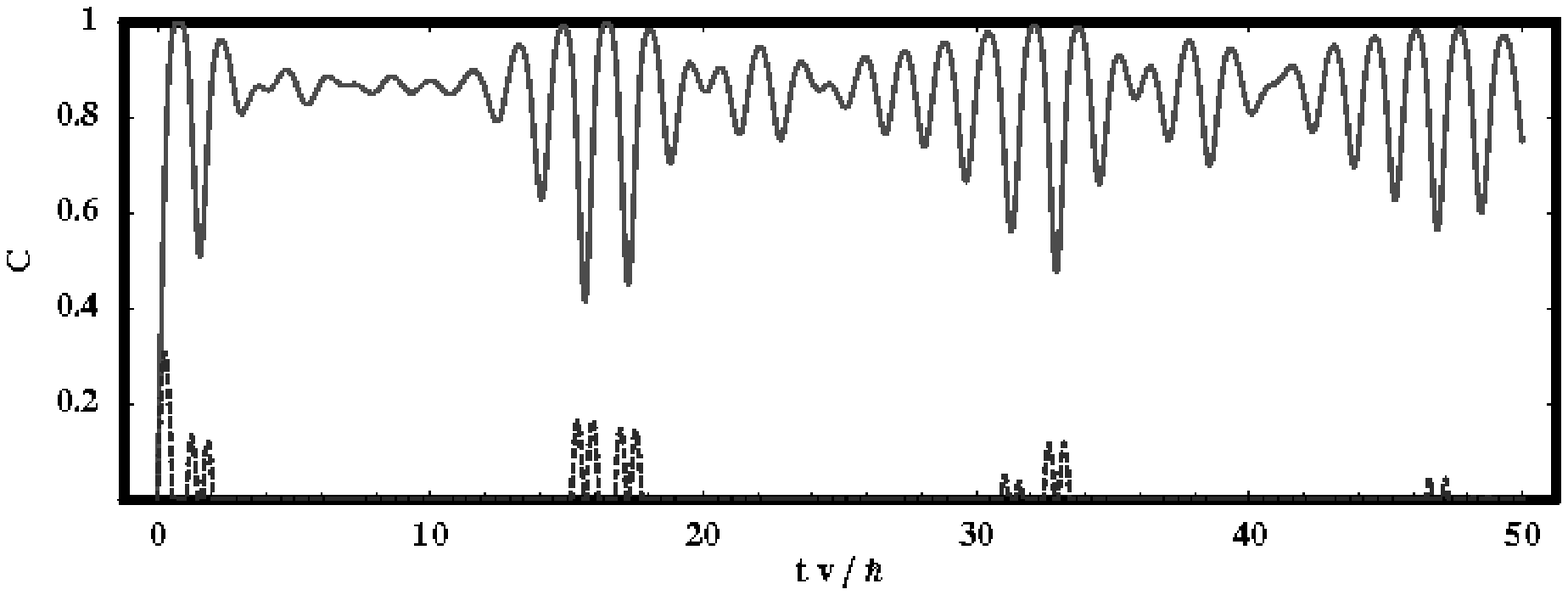}}}
    }
    \caption{\small Entrelazamiento de uno con el resto $C_1$ (línea roja sólida) y de a pares $C_2$ (línea azul a trazos) en función del tiempo en unidades de $\hbar/v$ con $b=0.5$, $g=1$ (saturan $C_1$ y $C_2$) para $n=5$ y $n=15$.}
    \label{fig:bogot.satura}
  \end{center}
\end{figure}
\begin{figure}[h]
  \begin{center}
    \mbox{
      \subfigure[n=5]{\scalebox{0.7}[0.7]{\epsfig{file=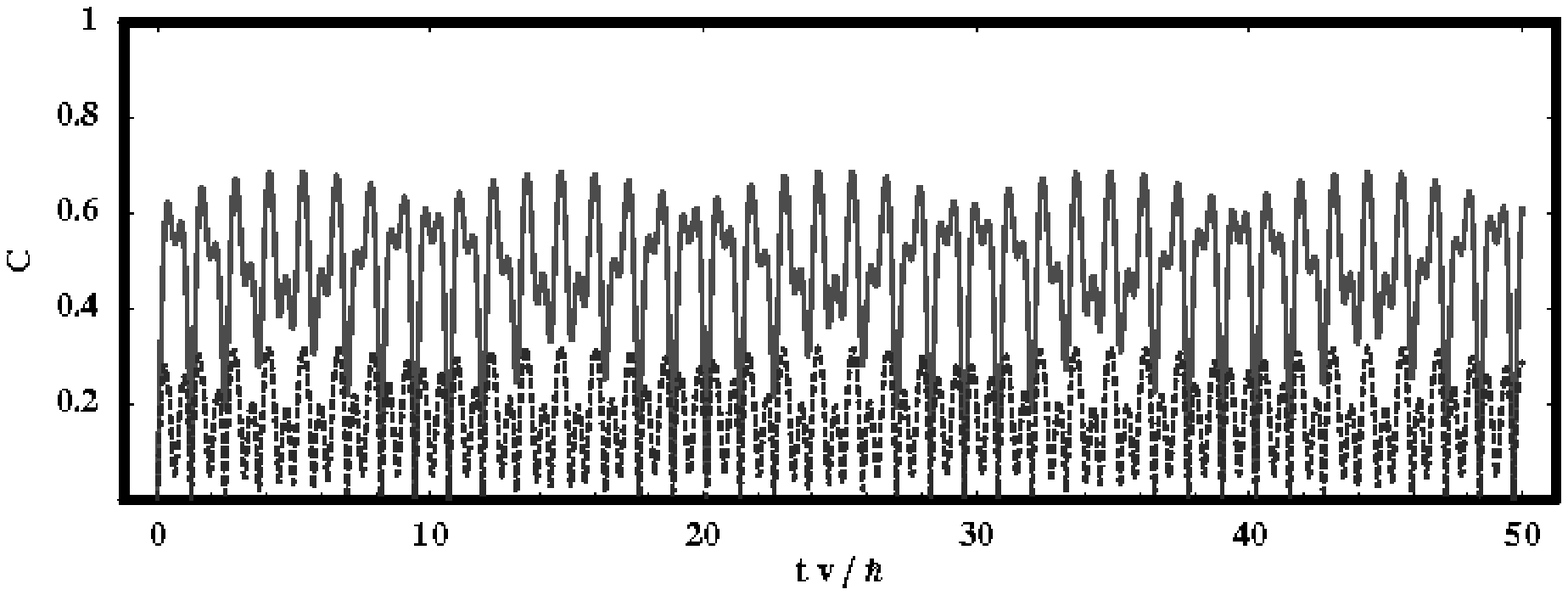}}}
    }
    \mbox{
      \subfigure[n=15]{\scalebox{0.7}[0.7]{\epsfig{file=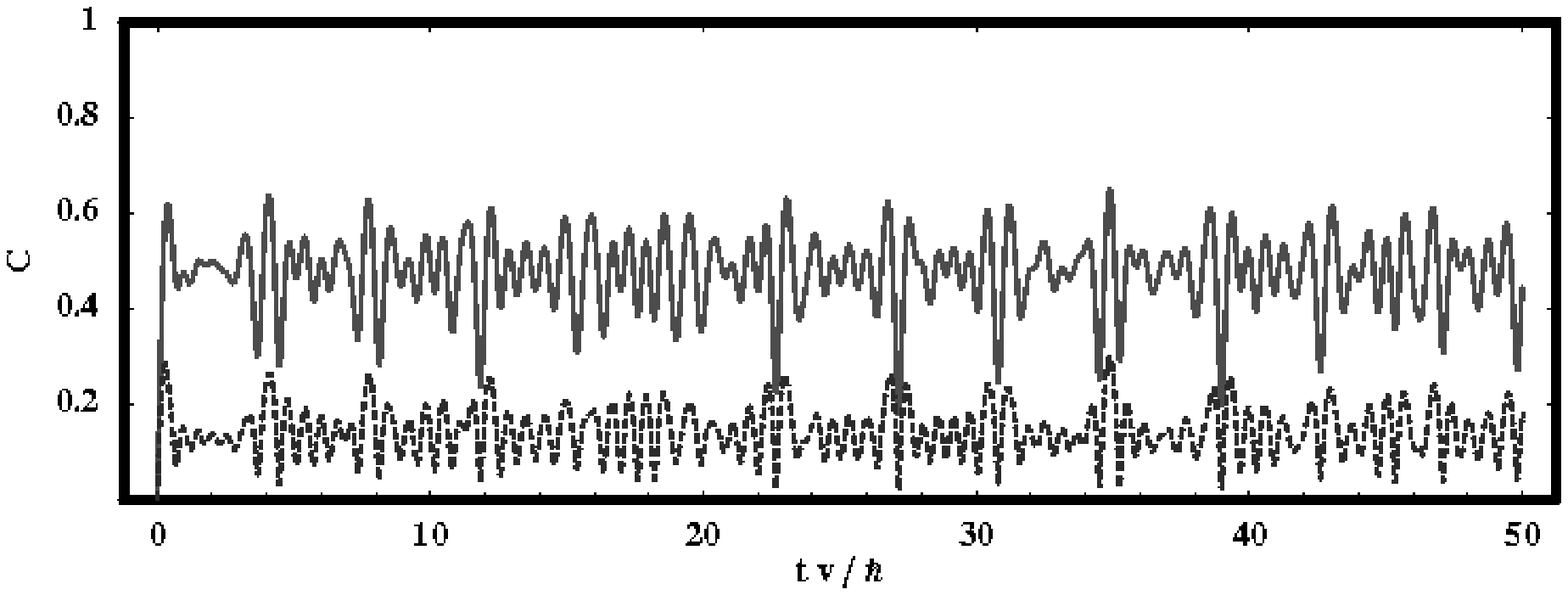}}}
    }
    \caption{\small Entrelazamiento de uno con el resto $C_1$ (línea roja sólida) y de a pares $C_2$ (línea azul a trazos) en función del tiempo en unidades de $\hbar/v$ con $b=4$, $g=1$ (no satura $C_1$ ni $C_2$) para $n=5$ y $n=15$.}
    \label{fig:bogot.nosatura}
  \end{center}
\end{figure}
La evolución temporal para sistemas de muchos espines es en general bastante complicada, pero hay casos para los que se pueden encontrar resultados relevantes, en los otros indicaremos generalidades.

Empezamos con los complicados. La figura \ref{fig:bogot.satura} muestra la evolución temporal del entrelazamiento en cadenas de 5 y 15 qubits para un campo magnético con el que se llega a la saturación $b=0.5$; $v=g=1$ (ver figs \ref{fig:bogon5.10} y \ref{fig:bogon15.10}). Se observa que para 5 qubits la dependencia temporal del entrelazamiento de uno con el resto es cuasiperiódica  y recorre continuamente todos los valores desde cero hasta la saturación mientras que para 15 qubits esta periodicidad parece desaparecer y el sistema no recorre todos los valores posibles del entrelazamiento de uno con el resto sino que oscila entre la saturación y una cota mínima. En realidad, ambos comportamientos son cuasiperiódicos y a un tiempo suficientemente largo es natural esperar que, para cualquiera de los casos, se de un momento en que el entrelazamiento de uno con el resto se anule. Sin embargo, para $n=15$ el entrelazamiento de uno con el resto permancece principalmente por arriba de una cota. El entrelazamiento de pares, en cambio, es menor y ocurre para tiempos más esporádicos cuando crece la cantidad de qubits en la cadena. 

El entrelazamiento de pares, de algún modo, es un indicador de que ``tan cuántico'' es un subsistema.  Al agrandarse la cadena la cantidad de espines que está monitoreando \footnote{Interactuando con, haciendoles de baño térmico u observando constantemente.} a los dos espines estudiados  se ve que la probabilidad de que se genere entrelazamiento en el subsistema es menor, es decir el sistema es ``menos cuántico''. Resumiendo, y a groso modo, se puede decir que, al agrandar el tamaño del sistema, un subsistema de un tamaño dado pierde la capacidad de que sus partes esten entrelazadas entre sí, pero estas permancecen constantemente entrelazadas con el resto del sistema\footnote{El resto del sistema viene a estar haciendo de medio ambiente o baño térmico para el subsitema que se está estudiando.} .  Esta observación es una de las ideas que dio a lugar a la teoría de la decoherencia que da un gran paso en la explicación de cómo, a pesar de que el mundo responde a las leyes e la mecánica cuántica, lo percibimos cómo clásico (ver apéndice \ref{ap.decoherencia}).

La figura \ref{fig:bogot.nosatura} muestra la evolución temporal del entrelazamiento para sistemas con la misma cantidad que qubits que antes pero con un campo magnético más fuerte que hace que el comportamiento este fuera de la saturación $b=4$; $v=g=1$ (ver figs \ref{fig:bogon5.10} y \ref{fig:bogon15.10}). El comportamiento, para el entrelazamieno de uno con el resto, es similar al anterior, pero en este caso el máximo alcanzado no llega a saturar. El entrelazamiento de pares también disminuye al agrandar la cadena, sin embargo, es mayor\footnote{Es mayor en el sentido de que, en promedio, hay más tiempos en el que el entrelazamiento de pares está por encima de una cota.} que en el caso anterior. El incremento en campo magnético disminuye el entrelazamiento global (de uno con el resto) impidiendo que las interacciones locales tengan infuencia a largo rango. Contrariamente se hacen más significativas las caracteristicas que solo dependen del corto rango. Para un campo magnético suficientemente grande el entrelazamiento de pares pácticamente ``sigue'' al entrelazamiento de uno con el resto\footnote{En estos casos de muchos qubits, sin embargo, aparecen comportamientos más complicados como saturaciones y ``revivals''.} del mismo modo que se había observado para el caso de solo tres espines (ver figura \ref{fig:ttresqbits5}).

\subsection{Resonancias}
\begin{figure}[t]
  \begin{center}
    \mbox{
      \subfigure[Entrelazamiento de uno con el resto $ C_1$.]{\scalebox{0.7}[0.7]{\epsfig{file=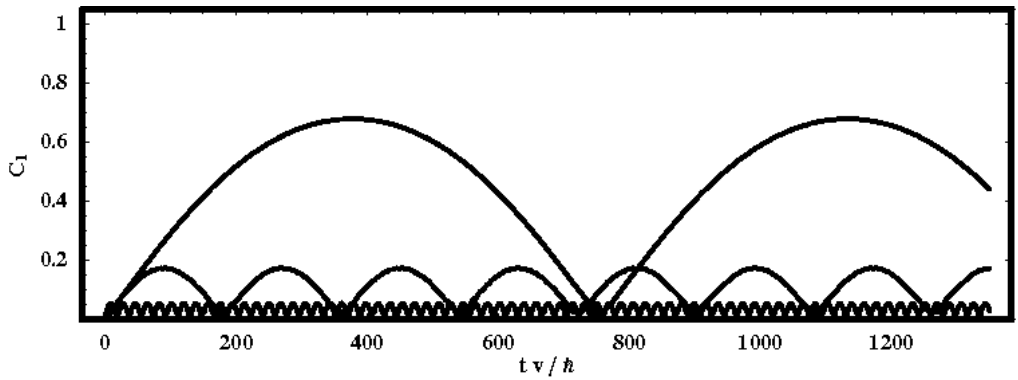}}\label{fig:bogot.resonancia}}
    }
    \mbox{
      \subfigure[Entrelazamiento de pares $C_2$.]{\scalebox{0.7}[0.7]{\epsfig{file=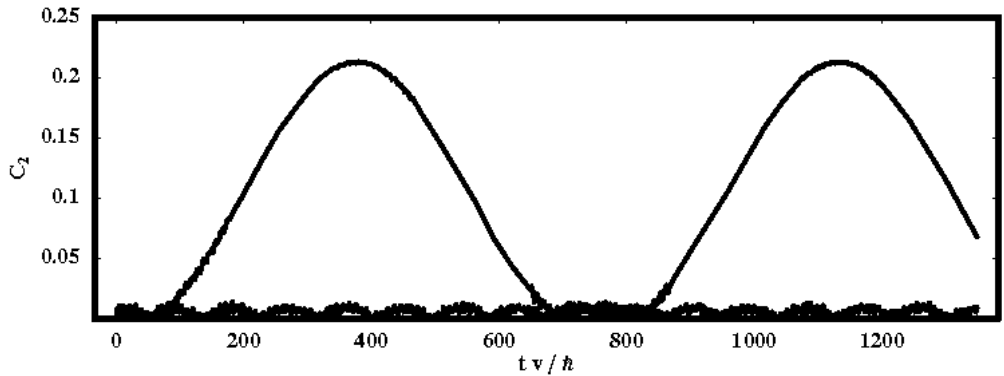}}\label{fig:bogot.resonancia-concu}}
    }
    \caption{\small Entrelazamientos $C_1$ y $C_2$ en función del tiempo en unidades de $\hbar/v$ en resonancia para el pico $j=7$ de un sistema de 15 qubits con $g=0.1$. (a) $C_1$ para campo resonante, a 1/10 de la distancia entre picos y justo entre picos. (b) $C_2$ para campo resonante y a 1/10 de la distancia entre picos (ver texto).}
      \end{center}
\end{figure}

En el caso en que hay resonancia, ver figuras \ref{fig:bogon5.01} y \ref{fig:bogon15.01}, la dependencia temporal es muy simple. La resonancia se da cuando la anisotropía de la interacción es muy chica $g\ll v$. Los campos para los cuales el sistema resuena cumplen $b=-2v\cos{w_j}$ de modo que los únicos términos relevantes en $a(t)$ son justamente el $j$ y el $n-j$, que corresponden ambos al mismo pico. En tal caso se puede aproximar $a(t)$ como:
\begin{eqnarray}
a(t)= \frac2n \sin^2{\lambda_jt}&;&\lambda_j=2g|\sin{w_j}|
\end{eqnarray}
La figura \ref{fig:bogot.resonancia} muestra la evolución temporal del entrelazamiento para un sistema de 15 qubits con $v=1$ y $g=0.1$. Se muestran resultados para el campo $b$ justo en la resonancia del pico de $j=7$, a 1/10 de la distancia entre este pico y el anterior y justo en el medio de estos dos picos \footnote{El pico $j=7$ es el pico de más a la derecha en la figura \ref{fig:bogon15.10}; el anterior $j=6$ es el segundo de derecha a izquierda.}. Es notable la sensibilidad a pequeñas variaciones de campo. Por más que en este caso el entrelazamiento no llegue a saturar, a los fines prácticos de construir una computadora cuántica este caso es de gran importancia ya que no sólo la dependencia temporal es simple sino que con pequeñas variaciones de campo puede ``prenderse'' o ``apagarse'' fácilmente la formación de entrelazamiento. También es importante notar que el único autovalor relevante del sistema $\lambda_j=2g|\sin{w_j}|$ se hace muy pequeño de modo que el tiempo necesario para llegar al máximo se hace muy grande en comparación con los tiempos naturales del sistema fuera de la resonancia. 

Cerca de la resonancia la concurrencia de pares para estados con paridad negativa tampoco es difícil de calcular. En particular se tiene que en este límite las ecuaciones \ref{eq:bogo.res} pueden aproximarse como:
\begin{eqnarray}
\braketo{S^+_{i}S^-_{i+1}}\approx
\frac2n |\cos{w_j}|\sin^2{\lambda_j t}\\   \braketo{S^+_{i}S^+_{i+1}}\approx 
-\frac1n|\sin{w_j}\sin{2\lambda_j t}|
\end{eqnarray}
de modo que los picos aparecen para algunos de los campos resonante en los tiempos en que $\sin{\lambda_j t}=1$ en los cuales la altura de estos es 
\begin{equation}
C_2=\frac4n\left[|\cos{w_j}|-|sin{w_j}|\sqrt{(1-\frac2n)^2-\frac2{n^2}\cos^2{w_j}}\right]
\end{equation}
Estos picos son de un valor mucho menor al del entrelazamiento de uno con el resto y aparecen adyacentes a los tiempos en los que este se anula y son de paridad opuesta al estado inicial. Se muestran en la figura \ref{fig:bogot.resonancia-concu} los resultados para el campo $b$ justo en la resonancia, a 1/10 de la distancia entre picos se ve que la sensibilidad del entrelazamiento de pares frente a pequeñas desviaciones de la resonancia es aún mayor que para el entrelazamiento de uno con el resto.\\

\subsection{Tiempos cortos}

\begin{figure}[h]
  \begin{center}
    \mbox{
      \subfigure[$n=5$] {\scalebox{0.5}{\epsfig{file=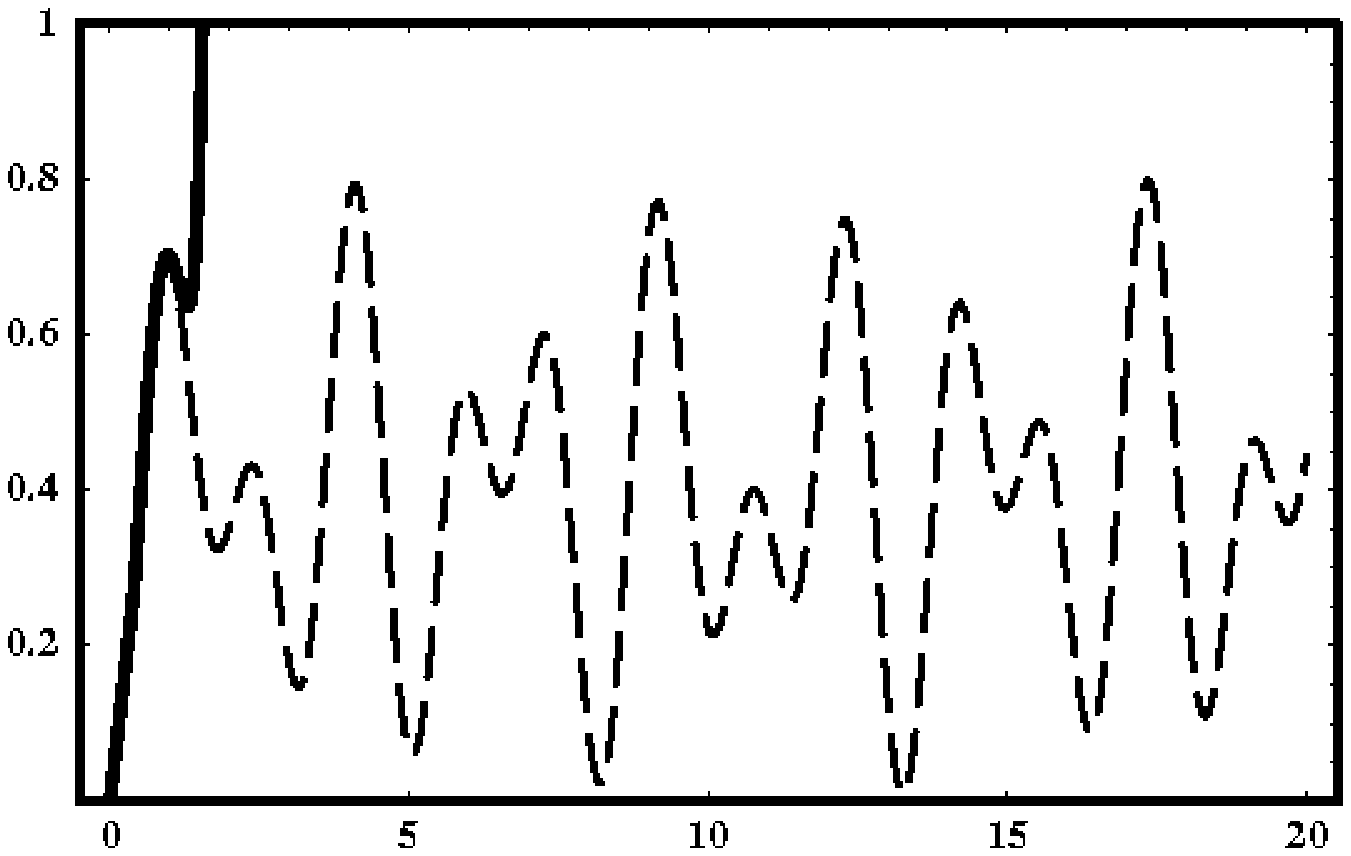}}}\quad
      \subfigure[$n=15$] {\scalebox{0.5}{\epsfig{file=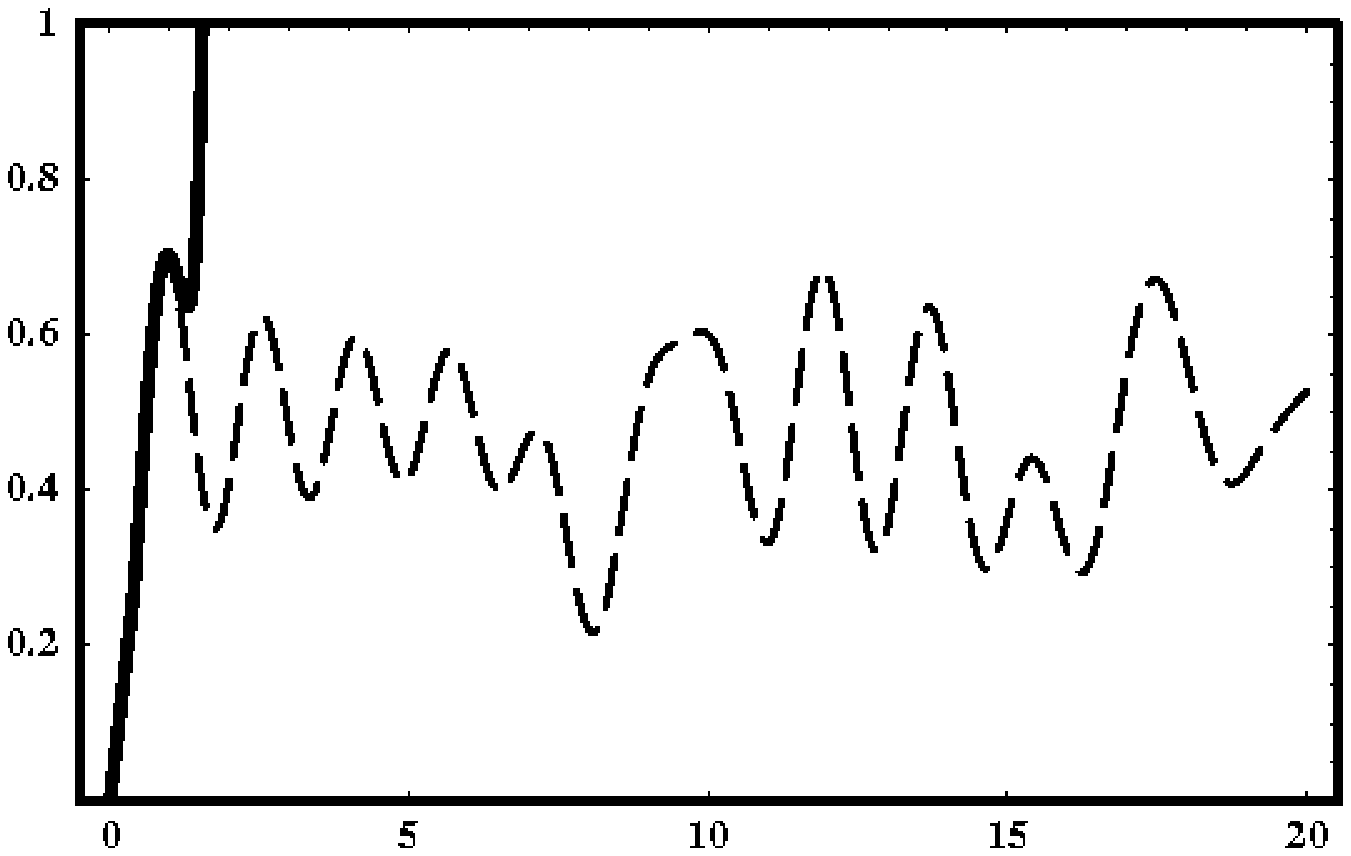}}}
    }
    \caption{\small Entrelazamientos de uno con el resto $C_1$ en función del tiempo en unidades de $\hbar/g$ y serie de Taylor en el caso que $b=v=0$ para $n=5$ y $n=15$.}
    \label{fig:bogot.approx}
  \end{center}
\end{figure}

Por último analizo el extremo $v=0$. En este caso todos lo picos colapsan en el origen. Para $b=0$ se tiene además que $a_m=(n-1)/n$ de modo que hay un tiempo para el cual se llega a la saturación. Lo interesante de este caso en particular es que es fácil hallar este valor de tiempo. El procedimiento consiste simplemente en hacer un desarrollo en serie de $a(t)$ de un orden suficientemente grande tal que siga a la función hasta que esta vale 1/2. El orden necesario del desarrollo para que cumpla esta condición es 10 (como puede verse en la fig \ref{fig:bogot.approx}) y el desarrollo que se obtiene, curiosamente, no depende de $n$.
\begin{equation}
a(t)\approx 2g^2t^2-2g^4t^4+\frac89 g^6t^6 -\frac29 g^8t^8 + \frac8{255} g^{10}t^{10} 
\end{equation}
El tiempo en el que se alcanza la saturación $a(t)=1/2$ es $t\approx0.6012/g$.

\chapter*{Conclusión}
\addcontentsline{toc}{chapter}{Conclusión}
Se examinó la formación de entrelazamiento a partir de un estado completamente alineado en una cadena XY con un campo magnético uniforme. Contrariamente a lo que se podría haber esperado, el comportamiento no es monótono con el campo magnético. El máximo entrelazamiento alcanzado presenta resonancias y ``plateaus''. 

Los ``plateaus'' ocurren para anisotropías que son del mismo orden o mayor que el parámetro de ``hopping'' y saturan cuando es mayor. Se mostró, además, que un aumento de la anisotropía no necesariamente garantiza una prolongación  del tiempo durante el cual el sistema permanece máximamente entrelazado. 

Las resonancias ocurren cuando la anisotropía es pequeña en comparación con el parámetro de ``hopping''. Son muy marcadas para el entrelazamiento de uno con el resto y más aun para el entrelazamiento de pares. Se encontró, además, que estas tienen un valor finito para anisotropías arbitrariamente chicas pero no nulas. Las resonancias permiten identificar configuraciones aptas para generar o no generar entrelazamiento en circuitos cuánticos con pequeñas variaciones del campo magnético.\\

Estos resultados y otros han sido aceptados para su publicación en la revista \emph{Physical Review A} en un trabajo conjunto con Raul Rossignoli. Entre los resultados que se incluyen figuran además: cadenas con cantidad par e impar de qubits, resonancia de pares entrelazados de paridad positiva, un estudio detallado del comportamiento para $v=g$ y $b=0$ donde la evolución es estrictamente periódica y resultados de diagonalización directa para $n=4$.

\part{Apéndices}
\appendix

\chapter{La descomposición de Schmidt}
\label{sec-schmidt}
El teorema de Schmidt es una herramienta muy poderosa para el análisis de sistemas bipartitos. Explico el teorema sin demostración\footnote{La demostración se basa en la Descomposición en Valores Singulares (DVS) y puede verse en el libro de Nielsen y Chuang \cite{NC}.}, menciono algunos corolarios inmediatos y hago un comentario.
\subsubsection{Teorema}
Sea un estado de un sistema bipartito de dimensión arbitraria descripto por un ket $\ket{\psi}$. Existen estados ortonormales $\ket{i_A}$ para el sistema $A$ y estados ortonormales $\ket{i_B}$ para el sistema $B$ tales que:
\begin{equation}
\ket{\psi}=\sum_i\lambda_i\ket{i_A}\ket{i_B}
\end{equation}
donde los $\lambda_i$ son números reales no negativos que satisfacen $\sum_i\lambda_i^2=1$ conocidos como \emph{coeficientes de Schmidt}. La cantidad de coeficientes de Schmidt no nulos se denomina \emph{número de Schmidt}.
\subsubsection{Corolarios}
Algunos corolarios que pueden demostrarse inmediatamente son:
\begin{itemize}
\item Los autovalores de las matrices reducidas de cada subsistema son iguales. 
\item Si un estado es separable su numero de Schmidt es 1.
\item El numero de Schmidt no cambia si se hacen operaciones unitarias sobre alguno de los subsistemas.
\end{itemize}
\subsubsection{Comentario}
A pesar de su potencia para revolver problemas la descomposición de Schmidt vale sólo para un par de subsistemas de dimensión arbitraria. Es decir, cada subsistema puede estar compuesto por más de un elemento pero no pueden siempre realizarse descomposiciones del tipo de Schmidt. Por ejemplo, para un sistema de tres componentes $A$,$B$,$C$ pueden realizarse descomposiciones del tipo $\sum_i\lambda_i\ket{i_{AB}}\ket{i_C}$, $\sum_i\lambda_i\ket{i_{A}}\ket{i_{BC}}$ o $\sum_i\lambda_i\ket{i_{AC}}\ket{i_B}$ donde los kets de cada subespacio son ortonormales y los $\lambda_i$ números reales positivos. En cambio no necesariamente pueden realizarse descomposiciones del tipo $\sum_i\lambda_i\ket{i_A}\ket{i_B}\ket{i_C}$ que cumplan que los kets de cada subespacio sean ortonormales y los $\lambda_i$ sean números reales no negativos.

\chapter{Paridad}
\label{chap-par}
\section{Paridad y Subsistemas}
Demostraré que si un estado tiene paridad definida, la matriz densidad reducida de cualquier subsistema conmuta con el operador paridad. $P\ket{\psi_{AB}}=\pm\ket{\psi}\Rightarrow\left[P,Tr_B\rho_{AB}\right]=0$.
\subsubsection{Definiciónes}
Defino el operador paridad general $P$ con autovectores $\ket{+}$,$\ket{-}$ de manera tal que $P\ket{+}=\ket{+}$ y $P\ket{-}=-\ket{-}$.\\
Un estado tiene paridad definida si es autoestado de $P$.
\subsubsection{Demostración}
En general podrá haber varios vectores ($\ket{+_i}$,$\ket{-_i}$) en una base que sean autovectores de $P$.
En particular, si estos vectores forman una base completa del espacio del Hilbert, cualquier estado podrá ser escrito como:
\begin{eqnarray}
\ket{\Psi}=\sum_{i}a_i \ket{+_i}+\sum_{i}b_i \ket{-_i}
\end{eqnarray}
Para un sistema bipartito esto se extiende a:
\begin{eqnarray}
\ket{\Psi_{AB}}=\sum_{ij}a_ib_j \ket{+_i}\ket{+_j}+\sum_{ij}g_ih_j \ket{-_i}\ket{-_j}\nonumber\\
\quad+\sum_{ij}c_id_j \ket{+_i}\ket{-_j}+
\sum_{ij}e_if_j \ket{-_i}\ket{+_j}
\end{eqnarray}
Es claro que los dos primeros términos tienen paridad positiva y los restantes negativa. Construyo entonces dos autovectores con paridad definida y veré si al tomar la traza parcial sobre un subsistema se conserva la paridad. Estos son (juntado las sumas):
\begin{eqnarray}
\ket{\Psi_{AB}^+}=\sum_{ijkl}\left\{a_ib_j \ket{+_i+_j}+g_kh_l\ket{-_k-_l}\right\}\nonumber\\
\ket{\Psi_{AB}^-}=\sum_{ijkl}\left\{c_id_j\ket{+_i-_j}+e_kf_l\ket{-_k+_l}\right\}
\end{eqnarray}
La matriz densidad para paridad positiva $\rho_{AB}^+\equiv\ket{\Psi_{AB}^+}\bra{\Psi_{AB}^+}$ es

\begin{eqnarray}
\rho_{AB}^+=
\sum_{ijklmnop}\left\{a_ib_ja_m^*b_n^*\ket{+_i+_j}\bra{+_m+_n}+g_kh_lg_o^*h_p^*\ket{-_k-_l}\bra{-_o-_p}\right.
\nonumber\\
\left.+a_ib_jg_o^*h_p^*\ket{+_i+_j}\bra{-_m-_n}+g_kh_la_m^*b_n^*\ket{-_k-_l}\bra{+_m+_n}\right\}
\end{eqnarray}
Al tomar la traza parcial sobre un subsistema ($B$) se obtiene la matriz reducida:
\begin{eqnarray}
\rho_{A}^+=\sum_{ijklmnop}\left\{
	a_ib_ja_m^*b_n^*\ket{+_i}\bra{+_m}+g_kh_lg_o^*h_p^*\ket{-_k}\bra{_o-}
\right\}\label{ec.rhoparidad}
\end{eqnarray}
Se ve claramente que $[\rho_{A}^+, P]=0$. La demostración para paridad negativa es exactamente igual. QED.\\
En realidad podría haberse demostrado lo anterior simplemente diciendo que cómo la traza parcial implica una reducción que deja solamente elementos que son diagonales en las partes trazeadas si la matriz densidad del sistema conmuta con $P$ tambien tiene que hacerlo la matriz reducida. Estos resultados también son válidos si $\rho_{AB}$ es un estado mixto.

\subsubsection{Caso $n=1$; un subsistema de dos niveles de distinta paridad}
En este caso, por ser 2 la dimensión del subsistema que se desea estudiar, la matriz reducida es la ecuación. \ref{ec.rhoparidad} pero sin las sumas.
\begin{eqnarray}
\rho_{A}^\pm=\ket{+}\bra{+}+\ket{-}\bra{-}\\
\end{eqnarray}
o bien en forma matricial (teniendo en cuenta que $Tr\rho_1=1$):
\begin{equation}
\rho_1=\left(\begin{array}{cc}a&0\\0&1-a\end{array}\right)
\end{equation}

\subsubsection{Caso $n=2$; dos subsistemas de dos niveles de distinta paridad}
En este caso la dimensión del subespacio es 4. Los autoestados de P son:
\begin{eqnarray}
\ket{+_i}&\rightarrow& \ket{++},\ket{--}\nonumber\\
\ket{-_i}&\rightarrow& \ket{+-},\ket{-+}
\end{eqnarray}
de modo que la matriz densidad será (directamente en forma matricial):
\begin{equation}
\rho_2=\left(\begin{array}{cccc}
a&0&0&d^*\\
0&b&e*&0\\
0&e&c&0\\
d&0&0&f\\
\end{array}\right)
\end{equation}
con $f=1-a-b-c$ debido a que $Tr\rho_2=1$.

\section{Paridad de espín}
\label{sec-parspines}
En un sistema de espines $1/2$ puede definirse un operador -\emph{paridad de espín}\footnote{En inglés se los suele llamar \emph{phase flip symmetry}.}-  que es un buen operador paridad asociado a la orientación de los espines. 
\begin{eqnarray}
P=e^{i\pi(\sum_i S^z_i + n/2)}=\prod_ie^{i\pi(S^z_i + n/2)}
\end{eqnarray}
donde $n$ es la cantidad de espines totales en el sistema.\\
Así definido este será un buen operador paridad definido sobre la base de $S^z$. Esto es la base $\ket{\su}$,$\ket{\sd}$ donde $S^z\ket{\su}=\frac12\ket{\su}$,$S^z\ket{\sd}=-\frac12\ket{\sd}$.\\
Se ve que éste es efectivamente un buen operador paridad estudiando como actúa sobre algunos estados. En particular sobre el estado $\ket{\su}^{\otimes n_\su}\ket{\sd}^{\otimes n_\sd}$ es inmeditado ver que:  
\begin{eqnarray}
P\ket{\su}^{\otimes n_\su}\ket{\sd}^{\otimes n_\sd}&=&
(-1)^{n_\su}\ket{\su}^{\otimes n_\su}\ket{\sd}^{\otimes n_\sd}
\label{ec:paridad}
\end{eqnarray}
Puesto en palabras: el operador paridad distingue si hay una cantidad par o impar de espines para ``arriba''.

\chapter{Transformada especial de Bogoliubov - BCS}
\label{ap-bogo.espe}
En este apéndice muestro como diagonalizar un Hamiltoniano específico utilizando una transformada especial de Bogoliubov, también llamada tranfsormada de BCS por Bardeen, Cooper y Shrieffer, que la usaron en su famoso paper de 1957 en el cual presentaron la primer teoría puramente cuántica de la superconductividad.

Sea un sistema de fermiones cuyos operadores de creación y aniquilación son $d_i$ y $d_i\da$ descripto por un Hamiltoniano del tipo
\begin{eqnarray}
H=\sum_{i=1}^n \left[  x_i \left(d_i\da d_i+d_{n-i}\da d_{n-i}\right)+ y_i\left(d_i\da d_{n-i}\da+d_{n-i}d_i\right)\right]
\end{eqnarray}
Un transformación especial de Bogoliubov de operadores de cuasipartículas fermiónicas $a_i$ y $a_i\da$ viene dada por 
\begin{eqnarray}
d_i\da=u_ia_i\da+v_ia_{n-i}\nonumber\\
d_{n-i}\da=u_ia_{n-i}\da-v_ia_{i}
\end{eqnarray}
donde se debe tener que $u_i$ y $v_i$ reales y $u_i^2+v_i^2=1$ para que los operadores de cuasipartículas sean fermiónicos. Haciendo los reemplazos correspondientes y  agrupando términos con iguales operadores se obtiene:
\begin{eqnarray}
H&=&\sum_{i=1}^n 
a_i\da a_i 
	\left[ x_i 	
		\left(
		u_i^2-v_i^2
		\right)
 	-2y_iu_iv_i
	\right] \nonumber\\
& &\hspace{1cm}+a_{n-i}\da a_{n-i}\left[ x_i 	
		\left(
		u_i^2-v_i^2
		\right)
 	-2y_iu_iv_i
	\right]  \nonumber\\
&&\hspace{1cm}+a_{n-i}\da a_{i}\da\left[ 
 	2x_iu_iv_i
		+y_i 	
		\left(
		u_i^2-v_i^2
		\right)
	\right]  \nonumber\\
&&\hspace{1cm}+a_{i} a_{n-i}\left[ 	
	2x_iu_iv_i
		+y_i 	
		\left(
		u_i^2-v_i^2
		\right)
	\right]  \nonumber\\
&&\hspace{1cm}+  -y_iu_iv_i+x_iv_i^2
\end{eqnarray}
Como se desea lleva el Hamiltoniano a una forma diagonal se debe exigir que $2x_iu_iv_i +y_i(u_i^2-v_i^2)$ se anule. Esta ecuación junto con la exigencia de que los operadores de cuasipartículas sean fermiones ($u_i^2+v_i^2=1$) tiene como posibles soluciones
\begin{eqnarray}
\left(\begin{array}{c}u\\v\end{array}\right)=
\pm\sqrt{\frac{\lambda_i\pm x_i}{2\lambda_i}}
&;&
\lambda_i=\sqrt{x_i^2+y_i^2}
\end{eqnarray}
de modo que el Hamiltoniano se escribe como:
\begin{eqnarray}
H=\sum_{i=1}^n \lambda_i\left(a_i\da a_i+a_{n-i}\da a_{n-i}\right) +\frac{x_i-\lambda_i}2
\end{eqnarray}

\chapter{Decoherencia}
\label{ap.decoherencia}
Con el fin de dar alguna idea sobre la decoherencia incluyo en este apéndice una traducción que hice de la introducción a un curso \cite{PZ2000} que dieron Paz y Zurek sobre Decoherencia inducida por el medio ambiente y la transición cuántico-clásica.

El origen cuántico del mundo clásico ha sido tan difícil de imaginar por los padres de la mecánica cuántica que llegaron a postular su existencia independiente, como lo hizo Niels Bohr, o trataron de buscar explicaciones más complejas con fundamentos clásicos, como lo hicieron de Broglie y Einstein. El problema radica en el principio cuántico de superposición que, en efecto, expande el posible espectro de estados posibles a una infinita superposición de ellos. Por ende superposiciones coherentes de gatos muertos y vivos tienen - dentro del formalismo cúantico - el mismo derecho de existir que cualquiera de sus dos alternativas clásicas (vivo o muerto). Dentro de todos los estados cuánticos posibles, los clásicamente permitidos son excepcionales. La cantidad de estados posibles para un sistema cuántico es enorme comparado con la cantidad que uno encuentra en sistemas clásicos. Sin embargo es un hecho verídico que sólo encontramos objetos clásicos en un subgrupo muy pequeño de todos los estados posibles (en principio permitidos). Debe uno explicar el origen de esta aparente regla de ``super-selección'' que previene la existencia de la mayoría de los estados posibles de un dado sistema. La \emph{decoherencia} y su principal consecuencia, la super-selección inducida por el medio ambiente -einselección-, dan cuenta de este hecho experimental del mundo físico.

La causa de la decoherencia es la interacción entre el sistema y su medio ambiente. Bajo una variedad de condiciones, que son muy fáciles de satisfacer para un objeto macroscópico, se da la einselección, de un pequeño subgrupo de estados cuasi-clásicos dentro de la enorme cantidad de estados cuánticos posibles. La clasicalidad es, entonces, una propiedad emergente inducida por el sistema y su interacción con el medio ambiente. Las superposiciones arbitrarias de estados permitidos son dejadas de lado en boga de un subgrupo preferido de ``estados punteros'' que emerge. Estos estados preferidos son los candidatos a los estados clásicos.  

El rol de los procesos de decoherencia en inducir la clasicalidad se ha aclarado sólo recién en la últimas tres décadas. La idea central es simple: El medio ambiente de un sistema cuántico puede, en efecto, monitorear sus estados mediante una continua interacción. La marca que los estados dejen en el medio ambiente contendrá la información de cuales serán los elegidos. Los estados que prevalecen son los que puedan interactuar con el medio ambiente sin ser perturbados. La propiedad clave que estos estados cuasi-clásicos tendrán es, entonces, su insensitividad al monitoreo y consecuentemente su resistencia al entrelazamiento causado por la interacción con el medio ambiente. Estos son también, casi por definición, los únicos estados que describen al sistema por sí sólo y no conjuntamente con el medio ambiente. Los otros estados son superposiciones entre el  sistema y medio ambiente y sólo conservan su pureza cuando ambos son analizados conjuntamente en un mayor ``super-sistema''.

El hecho que la interacción entre sistemas cuánticos produce entrelazamiento es bien conocido desde los principios de la mecánica cuántica. En efecto como las ideas de la decoherencia y la einselección se basan en la teoría cuántica, y solamente en la teoría cuántica, es lícito preguntarnos porque se tardó tanto en llegar a una explicación natural de los orígenes cuánticos de la clasicalidad. Hay varias posibles explicaciones para esta demora. El prejuicio que parece haber demorado más el estudio de sistemas cuánticos abiertos (que iteractuan con un medio ambiente) esta basado en un modo clásico de pensar sobre el Universo. Dentro del contexto de la física clásica todas las preguntas siempre se han podido responder pensando en un sistema cerrado (sin interacción con el medio ambiente). La estrategia básica consiste en asegurar la aislación del sistema involucrado agrandando el sistema a estudiar (incluyendo su medio ambiente más próximo). La expectativa era que de este modo uno podía reducir cualquier sistema abierto a uno más grande pero cerrado. Esta estrategia funciona en la física clásica donde el entrelazamiento puede ayudar a satisfacer leyes de conservación como la de la energía. Falla en el caso cuántico bajo consideración ya que en este caso es la información la que se desea que no se esparza. Es mucho más difícil contener información cuando un sistema se vuelve más grande. En fin, el único sistema macroscópico verdaderamente aislado es el Universo. Y nosotros, los observadores, ciertamente no estamos en condiciones de estudiarlo desde afuera.

\backmatter
\small
\bibliographystyle{alpha}
\bibliography{referencias}

\end{document}